\documentclass[twocolumn,preprintnumbers,amsfonts,amsmath,amssymb,nofootinbib,floatfix]{revtex4-1}
\usepackage{epsfig}
\usepackage{latexsym}
\usepackage{graphicx}% Include figure files
\usepackage{dcolumn}% Align table columns on decimal point
\usepackage{bm}% bold math

\renewcommand{\vec}[1]{\boldsymbol{#1}}
\newcommand{\be}{\begin{equation}}
\newcommand{\ee}{\end{equation}}
\newcommand{\bea}{\begin{eqnarray}}
\newcommand{\eea}{\end{eqnarray}}

\newcommand{\Tr}{\,\hbox{\rm Tr}}

\newcommand{\msbar}{{\rm \overline{MS\kern-0.05em}\kern0.05em}}

\newcommand{\<}{\langle}
\renewcommand{\>}{\rangle}
\newcommand{\ba}{\begin{eqnarray}}
\newcommand{\ea}{\end{eqnarray}}
\newcommand{\re}{\mathop{\rm Re}}
\newcommand{\im}{\mathop{\rm Im}}
\renewcommand{\vec}[1]{\boldsymbol{#1}}

\begin{document}

\title{Energy-momentum tensor on the lattice: non- perturbative 
renormalization in Yang--Mills theory}

\author{Leonardo Giusti$^{a,b}$ and Michele Pepe$^{b}$}

\affiliation{\vspace{0.3cm}
\vspace{0.1cm} 
$^a$ Dipartimento di Fisica, Universit\`a di Milano-Bicocca, 
Piazza della Scienza 3, I-20126 Milano, Italy\\
$^b$ INFN, Sezione di Milano--Bicocca,\\
Piazza della Scienza 3, I-20126 Milano, Italy\\
}

%\date{\vspace{0.2cm}}

\begin{abstract}
We construct an energy-momentum tensor on the lattice which 
satisfies the appropriate Ward Identities (WIs) and has the right trace anomaly in 
the continuum limit. It is defined by imposing suitable WIs associated to the 
Poincar\'e invariance of the continuum theory. These relations come forth 
when the length of the box in the temporal direction is finite, and they 
take a particularly simple form if the coordinate and the periodicity 
axes are not aligned. 
We implement the method for the SU(3) Yang--Mills theory discretized with 
the standard Wilson action in presence of shifted boundary conditions in the 
(short) temporal direction. By carrying out extensive numerical simulations, 
the renormalization constants of the traceless components of the tensor are 
determined with a precision of roughly half a percent for values of the 
bare coupling constant in the range $0\leq g^2_0\leq 1$.
\end{abstract}

\maketitle

\section{Introduction}
On the lattice the Poincar\'e group is explicitly broken 
into discrete subgroups, and the full symmetry is recovered 
only in the continuum limit. As a consequence, a given 
definition of the energy-momentum tensor needs to be 
properly renormalized to guarantee that the associated 
charges are the generators of the Poincar\'e group in 
the continuum limit, and that the trace anomaly is 
correctly reproduced.

In order to construct the renormalized energy-momentum tensor,
the way to proceed is to impose suitable WIs at fixed lattice
spacing that hold up to cutoff effects which vanish in the continuum 
limit~\cite{Caracciolo:1989pt}. This program can be 
realized in practice if the WIs involve correlators which 
in turn are simple enough to be computed numerically with 
good precision.

When the theory is considered in a finite box, the Euclidean Lorentz 
symmetry is also softly broken by its shape. If the length in one 
(temporal) direction $L_0$ is chosen to be shorter than the typical 
scale of the theory (thermal theory), interesting WIs 
follows~\cite{Giusti:2010bb,Giusti:2011kt,Giusti:2012yj}. 
They become particularly simple and of practical use if the periodicity
axes are tilted with respect to the lattice grid, i.e. if the hard 
breaking of the Poincar\'e symmetry due to the lattice discretization 
and the soft one due to the finite temporal direction are not aligned. 
This set-up has a natural implementation 
in the Euclidean path-integral formulation in terms of shifted 
boundary conditions~\cite{DellaMorte:2010yp,Giusti:2010bb}.

Here we define the renormalized energy-momentum tensor of the Yang--Mills theory 
non-perturbatively by working in this framework. This is achieved 
by supplementing the set of WIs found in Refs~\cite{Giusti:2011kt,Giusti:2012yj} 
with a new one which guarantees that the correct trace anomaly is reproduced 
in the continuum limit. 

We implement this strategy for the SU(3) Yang--Mills theory regularized with 
the standard Wilson action. We carry out extensive numerical simulations, and 
compute with high precision the renormalization constants of the traceless 
components of the tensor. The numerical determination of the renormalization constant 
of the trace part requires additional simulations, and it is left for a forthcoming 
publication. Over the last year an alternative method, based on the Yang--Mills 
gradient flow, has also been proposed for renormalizing non-perturbatively the 
energy-momentum tensor~\cite{Suzuki:2013gza,DelDebbio:2013zaa,Makino:2014taa}.  

\section{Ward identities in presence of a non-zero shift}
In this section we consider the SU(3) Yang--Mills theory in the continuum.
The definitions of the action and of the partition function $Z$
are reported in Appendices~\ref{app:appa} and \ref{app:appb} together with other conventions. 
Here we are interested in the thermal theory defined in the path integral 
formalism  with shifted boundary conditions 
\be\label{eq:shfbc}
A_\mu(L_0,\vec x) =A_\mu(0,\vec x - L_0\vec\xi)  
\ee
along the compact (temporal) direction of length $L_0$ with shift 
$\vec \xi \in\mathbb{R}^3$. The free-energy density is given by
\be\label{eq:freeE} 
f(L_0,\vec \xi) = - \frac{1}{L_0 V} 
\ln Z(L_0,\vec \xi)\; ,
\ee
where $V$ is the spatial volume of the box. In 
the thermodynamic limit, which is always assumed in this section, 
the invariance of the dynamics under the SO(4) group 
implies~\cite{Giusti:2012yj}
\be\label{eq:f}
f(L_0,\vec \xi) = f(L_0\sqrt{1+\vec \xi^2},\vec 0)\; .
\ee
When $\vec \xi \neq 0$, odd derivatives in $\xi_k$ do not vanish, and 
the following interesting relations
hold~\cite{Giusti:2010bb,Giusti:2011kt,Giusti:2012yj} ($x_0\neq 0$):
\ba\label{eq:dxi}
L_0 \langle \; T_{0k} \rangle_{\vec\xi} & = & \frac{1}{V}
\frac{\partial}{\partial \xi_k} \ln Z(L_0,\vec \xi)\;,\nonumber\\
\frac{\partial}{\partial \xi_k} \< O \>_{\vec\xi} & = & 
L_0 \langle \; {\overline T}_{0k}(x_0) \, O(0) \rangle_{\vec\xi,\, c}
\;,  
\ea
where $T_{\mu\nu}$ is the energy-momentum tensor, 
${\overline T}_{\mu\nu}=\int d^3x\, T_{\mu\nu}(x)$, $O$
is a generic gauge invariant operator, and the subscript $c$ 
indicates a connected correlation function.
By deriving once with respect to $L_0$ and to $\xi_k$
one obtains the relation (no summation over 
repeated $k$ here)~\cite{Giusti:2012yj}
\be\label{eq:WIodd}
\langle T_{0k} \rangle_{\vec\xi} = \frac{\xi_k}{1-\xi_k^2} 
\left\{\langle T_{00} \rangle_{\vec\xi}  
- \langle T_{kk} \rangle_{\vec \xi}\,\right\}\; .
\ee
In the equations above and in the rest of this paper we focus on
correlation functions of the energy-momentum tensor $T_{\mu\nu}$ with 
gauge-invariant operators inserted at a physical distance. 
As reviewed in Appendix~\ref{app:appc}, it is 
appropriate in those cases to consider the symmetric gauge-invariant 
definition of the energy-momentum tensor given by 
\be\label{eq:tmunuT}
T_{\mu\nu} = \frac{1}{g^2_0}\Big\{F^a_{\mu\alpha}F^a_{\nu\alpha} 
- \frac{1}{4}\delta_{\mu\nu}\, F^a_{\alpha\beta}F^a_{\alpha\beta}\Big\}\; . 
\ee
By deriving two times with respect to the shift components
and by using Eq.~(\ref{eq:dxi}), one obtains~\cite{Giusti:2012yj}
\be
\label{eq:wipv}
\langle T_{0k}\rangle_{\vec\xi}  = 
\frac{L_0 \xi_k}{2}\,  \sum_{ij}\,
\left\langle {\overline T}_{0i}\,  T_{0j} \right\rangle_{\vec\xi,\, c}\, 
\left[ \delta_{ij} - \frac{\xi_i\, \xi_j}{\vec\xi^2}\right] 
\ee
where on the r.h.s. the two fields are inserted at different times.
Analogously one shows that ($x_0 \neq 0$)
\ba\label{eq:sing}
& & L_0 \langle \; {\overline T}_{0k}(x_0) \, T_{\mu\mu}(0) \rangle_{\vec\xi,\, c}
= \left\{6 - \frac{1+\xi^2}{\xi_k^2} \right\}  
\langle \; T_{0k} \rangle_{\vec\xi} +\nonumber\\ 
& & \qquad\qquad\qquad\qquad  
L_0 \frac{1+\xi^2}{\xi_k} \langle \; {\overline T}_{0k}(x_0) 
\, T_{0k}(0) \rangle_{\vec\xi,\, c}\; ,
\ea 
which can also be put in the more suggestive form 
\be\label{eq:sing2}
\frac{\partial}{\partial \xi_k} \langle T_{\mu\mu} \rangle_{\vec\xi} =
\frac{1}{(1+\xi^2)^2}\frac{\partial}{\partial \xi_k}
\left[\frac{(1+\xi^2)^3}{\xi_k} \langle T_{0k} \rangle_{\vec\xi} \right]\; . 
\ee

\subsection{Finiteness of $T_{\mu\nu}$ and trace anomaly}
In dimensional regularization, the energy-momentum tensor in 
Eq.~(\ref{eq:tmunuT}) is decomposed as
\be
T_{\mu\nu} = \tau_{\mu\nu} + \delta_{\mu\nu} \tau \; , 
\ee
where 
\ba
\tau_{\mu\nu} & = & \frac{1}{g_0^2}
\left\{F^a_{\mu\alpha} F^a_{\nu\alpha} - 
\frac{1}{D}\delta_{\mu\nu} F^a_{\alpha\beta} F^a_{\alpha\beta} \right\}\; , \\ 
\tau & = & \frac{\epsilon}{2 D g_0^2} F^a_{\alpha\beta} F^a_{\alpha\beta}\nonumber
\ea
are two fields transforming as a two-index symmetric and a singlet irreducible representation of the SO(D) group
respectively, and $D=4-2\epsilon$.

The field  $\tau_{\mu\nu}$ is a dimension-four gauge invariant operator which 
is multiplicatively renormalizable. The WI in Eq.~(\ref{eq:wipv}) fixes its renormalization constant to 1. 
This in turn implies that $g_0^2\, \tau_{\mu\nu}$ renormalizes as 
\ba
\left\{F^a_{\mu\alpha} F^a_{\nu\alpha} - \frac{1}{D} \delta_{\mu\nu} F^a_{\alpha\beta} 
F^a_{\alpha\beta}\right\}^{\rm R} & = & Z_g 
\Big\{F^a_{\mu\alpha} F^a_{\nu\alpha}\\
 & & \left. - \frac{1}{D} \delta_{\mu\nu} F^a_{\alpha\beta} 
F^a_{\alpha\beta}\right\}\; ,\nonumber 
\ea
where $Z_g$ is the renormalization constant of the coupling, see Appendix~\ref{app:appd}.
By defining the renormalization group invariant (RGI) operator as~\cite{Luscher:2013lgggg} 
\ba
\left\{F^a_{\mu\alpha} F^a_{\nu\alpha} - \frac{1}{D} \delta_{\mu\nu} F^a_{\alpha\beta} 
F^a_{\alpha\beta}\right\}^{\rm RGI} & = & \frac{1}{2 b_0 g^2}
\Big\{F^a_{\mu\alpha} F^a_{\nu\alpha}\\ 
&  & \left. - \frac{1}{D} \delta_{\mu\nu} F^a_{\alpha\beta} 
F^a_{\alpha\beta}\right\}^{\rm R}\; ,\nonumber 
\ea
we finally arrive to
\be
\tau_{\mu\nu} = \tau^{\rm R}_{\mu\nu} = 
2 b_0 \left\{F^a_{\mu\alpha} F^a_{\nu\alpha} - \frac{1}{D} \delta_{\mu\nu} F^a_{\alpha\beta} 
F^a_{\alpha\beta}\right\}^{\rm RGI},  
\ee
where $b_0$ is the first coefficient of the $\beta$-function given in 
Eq.~(\ref{eq:bo1}).
The field $\tau$ is also dimension-four and gauge invariant, but it is a singlet 
under SO(D). Therefore it mixes with itself and with the identity operator. The 
Eq.~(\ref{eq:sing}) fixes the multiplicative renormalization constant to 1, while 
a natural prescription for the identity subtraction is 
\be\label{eq:tau1}
\tau^{\rm R}  = \tau - \langle \tau \rangle_0
\ee
where $\langle \dots \rangle_0$ indicates the vacuum expectation value 
for $L_0\rightarrow \infty$ (zero temperature). This in turn implies that
one can define
\be\label{eq:ZE}
\left\{F^a_{\alpha\beta} F^a_{\alpha\beta}\right\}^{\rm R} = Z^{-1}_E 
\left\{F^a_{\alpha\beta} F^a_{\alpha\beta} - \langle F^a_{\alpha\beta} F^a_{\alpha\beta}\rangle_0 \right\}\; ,
\ee
and Eq.~(\ref{eq:tau1}) becomes 
\be
\tau^{\rm R} = \frac{\epsilon Z_g Z_E}{2D \mu^{2\epsilon} g^2} \left\{F^a_{\alpha\beta} F^a_{\alpha\beta}\right\}^{\rm R}\; . 
\ee
By using the result in Eq.~(\ref{eq:ZEr}) of Appendix \ref{app:appd}, one obtains
the well known result for the trace anomaly~\cite{Adler:1976zt,Collins:1976yq}
\be\label{eq:Tran}
T_{\mu\mu} = 4 \tau^{\rm R} = -\frac{\beta}{2 g^3} 
\left\{F^a_{\alpha\beta} F^a_{\alpha\beta}\right\}^{\rm R}\; .
\ee
By defining the renormalization group invariant operator as~\cite{Luscher:2013lgggg} 
\be
\{F^a_{\alpha\beta}F^a_{\alpha\beta}\}^{\rm RGI} = - \frac{\beta}{b_0 g^3}
\{F^a_{\alpha\beta}F^a_{\alpha\beta}\}^R\; ,
\ee
we can finally write
\be\label{eq:Tran2}
T_{\mu\mu} = 4 \tau^{\rm R} = 
\frac{b_0}{2} \{F^a_{\alpha\beta}F^a_{\alpha\beta}\}^{\rm RGI}\; . 
\ee
The WIs in Eqs.~(\ref{eq:wipv}) and (\ref{eq:sing}) fix unambiguously 
the renormalization constants of the composite fields entering the 
energy-momentum tensor definition so that the correct trace anomaly is 
reproduced. The Eqs.~(\ref{eq:Tran})--(\ref{eq:Tran2}) 
hold to all orders in perturbation theory.

\section{The energy momentum tensor on the lattice}
We regularize the SU(3) Yang--Mills theory on a finite 
four-dimensional lattice of spatial volume $V=L^3$,
temporal direction $L_0$, and spacing $a$. The gauge
field satisfies periodic boundary conditions in the 
three spatial directions and shifted boundary conditions 
in the compact direction 
\be
U_\mu(L_0,\vec x) =U_\mu(0,\vec x - L_0\vec\xi)\; ,
\ee
where $U_\mu(x_0,\vec x)$ are the link variables.
The action is discretized through the standard Wilson 
plaquette 
\be\label{eq:latS}
S[U] = \frac{\beta}{2}\, a^4\sum_{x} \sum_{\mu,\nu} 
\left[1 - \frac{1}{3}{\rm Re}\Tr\Big\{U_{\mu\nu}(x)\Big\}\right]\; ,
\ee
where the trace is over the color index, and $\beta=6/g_0^2$ with $g_0$ 
being the bare coupling constant. The plaquette is defined as a function 
of the gauge links, and it given by
\be\label{eq:placst}
U_{\mu\nu}(x) = U_\mu(x)\, U_\nu(x+ a\hat \mu)\, U^\dagger_\mu(x + a\hat \nu)\,
             U^\dagger_\nu(x)\; ,  
\ee
where $\mu,\,\nu=0,\dots,3$, $\hat \mu$ is the unit vector along the 
direction $\mu$, and $x$ is the space-time coordinate. The gluon field strength 
tensor is defined as\footnote{We use the same
notation for lattice and continuum quantities, since any ambiguity is 
resolved from the context. As usual, the continuum limit value of a
renormalized lattice quantity, identified with the subscript ${\rm R}$, is the one to 
be identified with its continuum counterpart.}~\cite{Caracciolo:1989pt} 
\be
F^a_{\mu\nu}(x) = - \frac{i}{4 a^2} 
\Tr\Big\{\Big[Q_{\mu\nu}(x) - Q_{\nu\mu}(x)\Big]T^a\Big\}\; , 
\ee
where 
\ba
& & Q_{\mu\nu}(x) =  
U_\mu(x)\, U_\nu(x+ a\hat\mu)\, U^\dagger_\mu(x + a\hat\nu)\,U^\dagger_\nu(x)\nonumber\\[0.125cm]
& & + U_\nu(x)\, U^\dagger_\mu(x-a\hat\mu+a\hat\nu)\, U^\dagger_\nu(x - a\hat\mu)\,
U_\mu(x-a\hat\mu)\\[0.125cm]
& & + U^\dagger_\mu(x-a\hat\mu)\, U^\dagger_\nu(x-a\hat\mu-a\hat\nu)\, 
U_\mu(x-a\hat\mu-a\hat\nu)\,U_\nu(x-a\hat\nu)\nonumber\\[0.125cm]
& & + 
U^\dagger_\nu(x-a\hat\nu)\, U_\mu(x-a\hat\nu)\, 
U_\nu(x + a\hat\mu - a\hat\nu)\, U^\dagger_\mu(x)\; . \nonumber
\ea
The target energy-momentum tensor in the continuum
is a gauge-invariant operator of dimension 4, which is a 
combination of a traceless two-index symmetric and a singlet 
irreducible representation of SO(4). When SO(4) is broken
to the hypercubic group SW$_4$, the traceless two-index symmetric representation
splits into a sextet (non-diagonal components) and a triplet (diagonal
traceless components). At finite lattice spacing, the energy-momentum
tensor is thus a combination of gauge-invariant operators of dimension $d\leq 4$ 
which, under the hypercubic group, transform as 
one of those two representations and the singlet. In the SU(3) Yang--Mills theory 
there are only three such operators 
(no summation over repeated $\mu$ and $\nu$ here)~\cite{Caracciolo:1989pt,Caracciolo:1991cp}:
\ba\label{eq:lattmunu}
T^{[1]}_{\mu\nu} & = & (1-\delta_{\mu\nu}) \frac{1}{g_0^2} \Big\{F^a_{\mu\alpha}F^a_{\nu\alpha} \Big\}\nonumber\\[0.125cm]
T^{[2]}_{\mu\nu} & = & \delta_{\mu\nu}\, \frac{1}{4 g_0^2}\, F^a_{\alpha\beta} F^a_{\alpha\beta}\\[0.125cm]
T^{[3]}_{\mu\nu} & = & \delta_{\mu\nu} \frac{1}{g_0^2} \Big\{F^a_{\mu\alpha}F^a_{\mu\alpha} 
- \frac{1}{4} F^a_{\alpha\beta}F^a_{\alpha\beta} \Big\}\nonumber
\ea
and the identity. The sextet $T^{[1]}_{\mu\nu}$ and the triplet $T^{[3]}_{\mu\nu}$ renormalize 
multiplicatively, while the singlet $T^{[2]}_{\mu\nu}$ mixes also with the identity.
The renormalized energy-momentum tensor can finally be written as
\be
T^{\rm R}_{\mu\nu} = Z_{_T} \Big\{T^{[1]}_{\mu\nu} + z_{_T}  T^{[3]}_{\mu\nu} + 
z_{_S} \big[T^{[2]}_{\mu\nu} - 
\langle T^{[2]}_{\mu\nu} \rangle_0 \big]\Big\}\; .
\ee
The renormalization constants $Z_{_T}$, $z_{_T}$ and $z_{_S}$ are finite 
and depend on $g^2_0$ only. At one loop
in perturbation theory their expressions are~\cite{Caracciolo:1989pt,Caracciolo:1991cp}
\ba\label{eq:PT1loop}
Z_{_T}(g_0^2) & = & 1 + 0.27076 \; g_0^2\;, \nonumber\\
z_{_T}(g_0^2) & = & 1 - 0.03008\; g_0^2 \;, \\ 
z_{_S}(g_0^2) & = & \frac{b_0}{2} g_0^2\; . \nonumber 
\ea
\begin{figure*}[t]
\includegraphics[width=6.7 cm,angle=0]{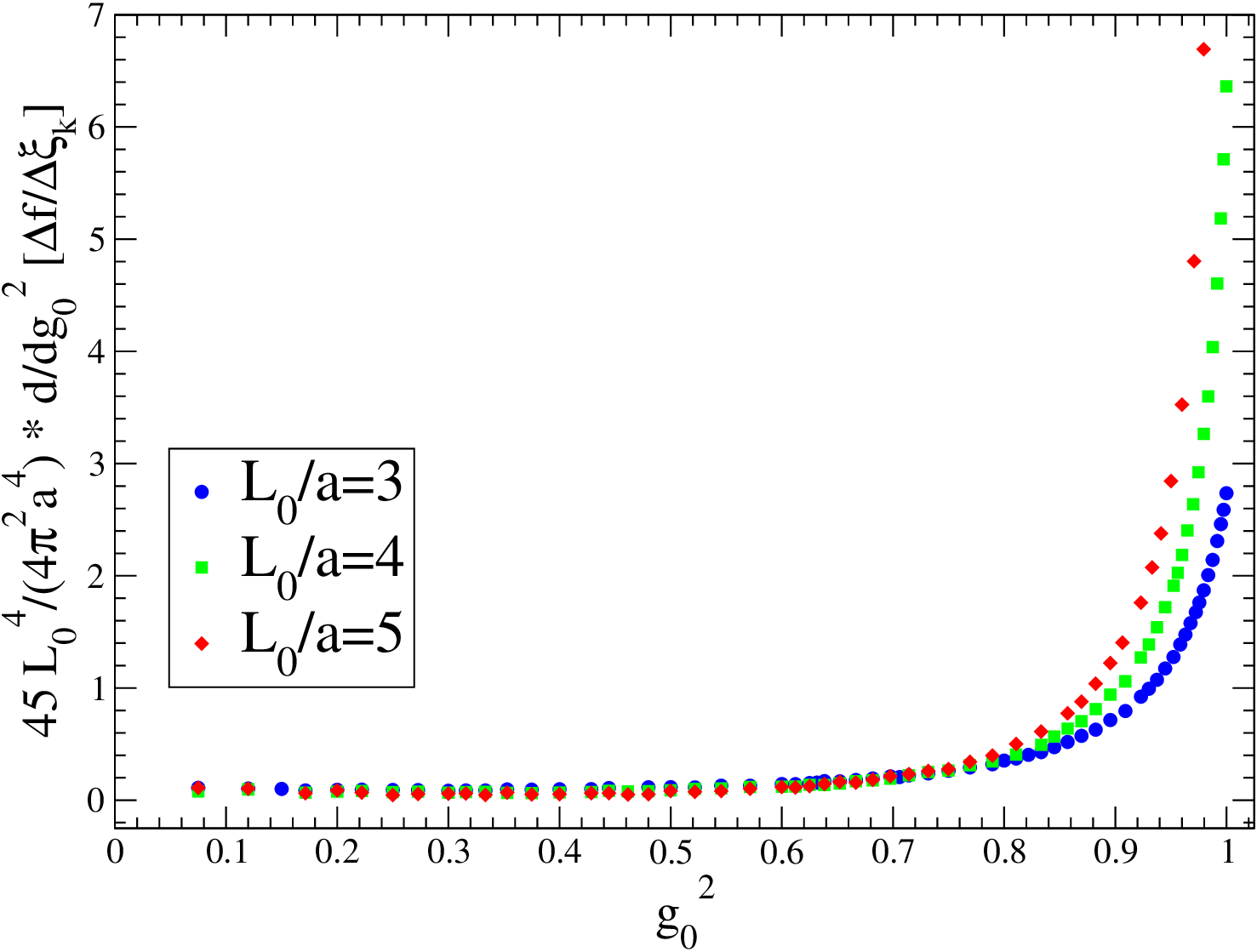}\ \ \ \ \
\includegraphics[width=7.0 cm,angle=0]{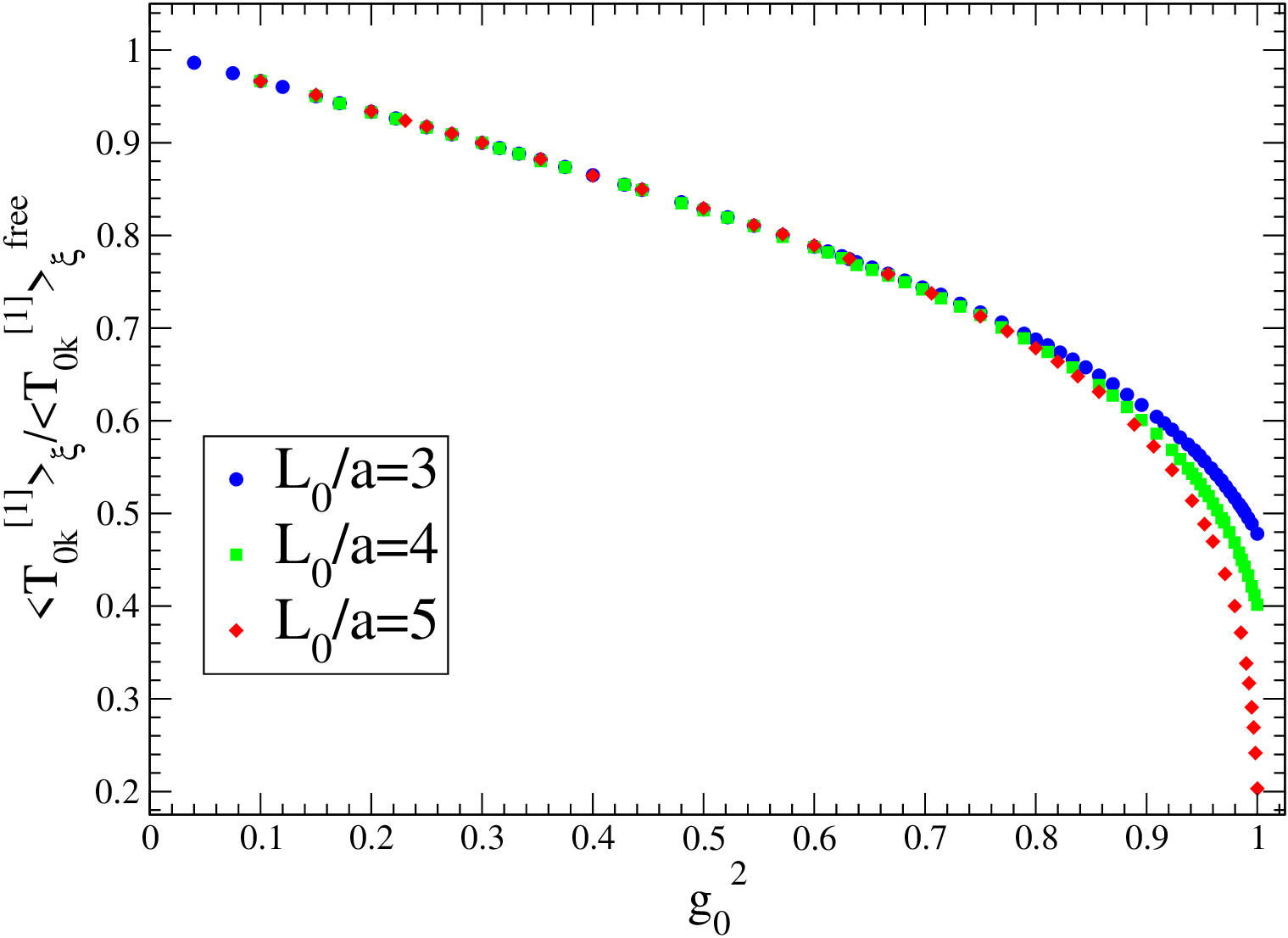}
\caption{Left: the derivative of $\Delta f/\Delta \xi_k$ (normalized to 
its Stefan--Boltzmann value) with respect to $g_0^2$ 
as a function of the bare coupling. Right:
$\langle T^{[1]}_{0k} \rangle_{\vec\xi}$ normalized to its tree-level expression
as a function of $g_0^2$. The data are generated on lattices with $L_0/a=3$ (blue), 
$4$ (green) and $5$ (red), $L/a=48$, and $\vec\xi=(1,0,0)$. Statistical 
errors are smaller than symbols.\label{fig:bare_res}}
\end{figure*}

\subsection{Non-perturbative renormalization conditions\label{eq:secNP}}
The renormalization constants $Z_{_T}$, $z_{_T}$ and $z_{_S}$
can be determined non-perturbatively by requiring that on the 
lattice the WIs in Eqs.~(\ref{eq:dxi}), (\ref{eq:WIodd}), 
and (\ref{eq:sing2}) hold up to discretization 
effects which vanish in the continuum limit. The renormalization 
constant of the sextet is fixed to 
be~\cite{Giusti:2014ila} (see also~\cite{Robaina:2013zmb}) 
\be\label{ZTpractic}
Z_{_T}(g_0^2) = - \frac{\Delta f}{\Delta \xi_k}\, \frac{1}{\langle T^{[1]}_{0k} \rangle_{\vec\xi} }\; , 
\ee
where the derivative in the shift in Eq.~(\ref{eq:dxi}) is discretized
by the symmetric finite difference 
\be\label{eq:discrder}
\frac{\Delta f}{\Delta \xi_k} = \frac{1}{2a V}
\ln\Big[\frac{Z(L_0,{\vec\xi }-a \hat k/L_0 )}{Z(L_0,{\vec\xi}+a \hat k/L_0 )} \Big] 
\ee
to ensure that discretization effects start at $O(a^2)$.
In the thermodynamic limit, which is always assumed in this section, the 
triplet is renormalized by requiring that Eq.~(\ref{eq:WIodd}) holds up 
to harmless discretization effects, i.e.
\begin{equation}\label{Zd}
z_{_T}(g_0^2) = \frac{1-\xi_k^2}{\xi_k} \frac {\langle T^{[1]}_{0k}
  \rangle_{\vec\xi} }{\langle T^{[3]}_{00}\rangle_{\vec\xi} - \langle T^{[3]}_{kk} \rangle_{\vec\xi}}\; . 
\end{equation}
By choosing one possibility of discretizing Eq.~(\ref{eq:sing2}), the singlet renormalization 
constant is fixed to be
\begin{widetext}
\be\displaystyle
z_{_S} = \frac{1}{(1+\xi^2)^2}\frac{\left[\frac{(1+\xi^{'2})^3}{\xi'_k} \langle T^{[1]}_{0k} \rangle_{\vec\xi'} 
\right]_{\vec\xi'=\vec\xi+ a \hat k/L_0} - 
\left[\frac{(1+\xi^{'2})^3}{\xi'_k} \langle T^{[1]}_{0k} \rangle_{\vec\xi'} 
\right]_{\vec\xi'=\vec\xi- a \hat k/L_0}
}
{\langle T^{[2]}_{\mu\mu} \rangle_{\vec\xi+ a \hat k/L_0} - \langle T^{[2]}_{\mu\mu} \rangle_{\vec\xi- a \hat k/L_0}}\; . 
\ee
\end{widetext}
At finite $L_0$, the renormalization constants depend on the bare 
coupling constant and on $(a/L_0)^2$ due to discretization effects. 
Our prescription is to define them in 
the limit\footnote{Notice that in Ref.~\cite{Giusti:2014ila} a 
different condition was imposed. Since we were interested in $Z_{_T}(g_0^2)$ in
a limited range of $g_0^2$, we defined $Z_{_T}(g_0^2)$ as in Eq.~(\ref{ZTpractic}) 
but at finite $L_0$.} $L_0\rightarrow \infty$ at fixed $g^2_0$.

\section{Numerical computation}
In this section we describe how the strategy outlined above
has been implemented in practice to determine the renormalization 
constants $Z_{_T}$ and $z_{_T}$. In all simulations the 
basic Monte Carlo step is a combination of heatbath and 
over-relaxation updates of the link variables using the Cabibbo--Marinari 
scheme~\cite{Cabibbo:1982zn}. A single sweep is made 
of 1 heatbath and 3 over-relaxation updates 
of all link variables. All lattices have an inverse temporal length $1/L_0>T_c$, 
where $T_c$ is the critical temperature of the theory. We have checked 
explicitly the autocorrelation times of the primary observables by profiting 
from the long Monte Carlo histories, which are typically made of $O(10^5) $ sweeps. 
No long autocorrelations were observed\footnote{At these values of $L_0$
fluctuations of the topological charge away from zero are heavily 
suppressed.}. For the statistical analysis we have blocked together 
the primary observables generated in several hundreds of consecutive sweeps, 
a value which is always much larger than the autocorrelation times measured.
It is important to notice that the determinations of $Z_{_T}$  and $z_{_T}$ require (see below) 
the  computation of expectation values of single local operators only. Indeed increasing the 
spatial size of the lattice does not increase the computational effort 
at fixed statistical accuracy.
\begin{figure*}[t]
\includegraphics[width=7.0 cm,angle=0]{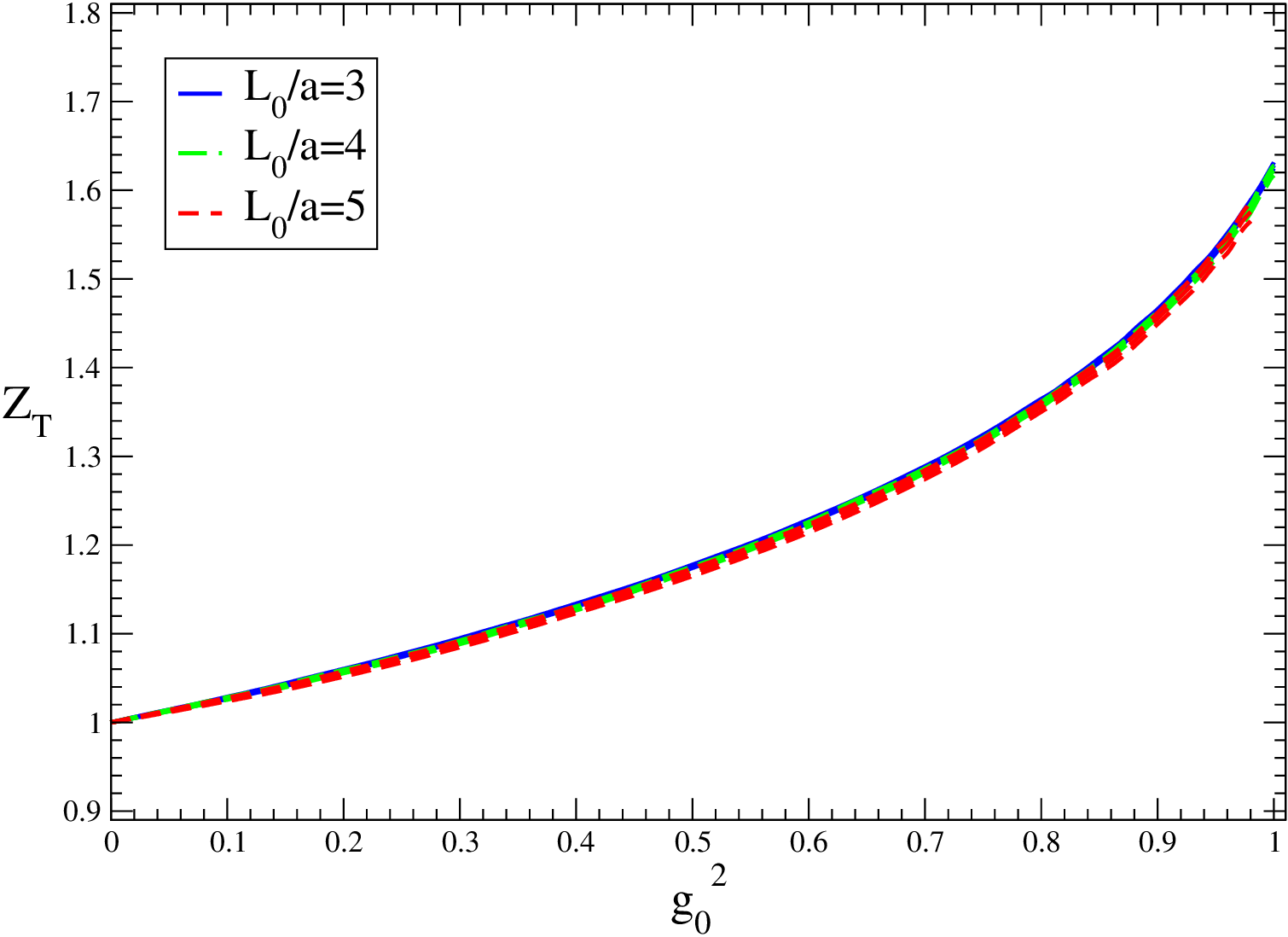}\ \ \ \ \ 
\includegraphics[width=7.0 cm,angle=0]{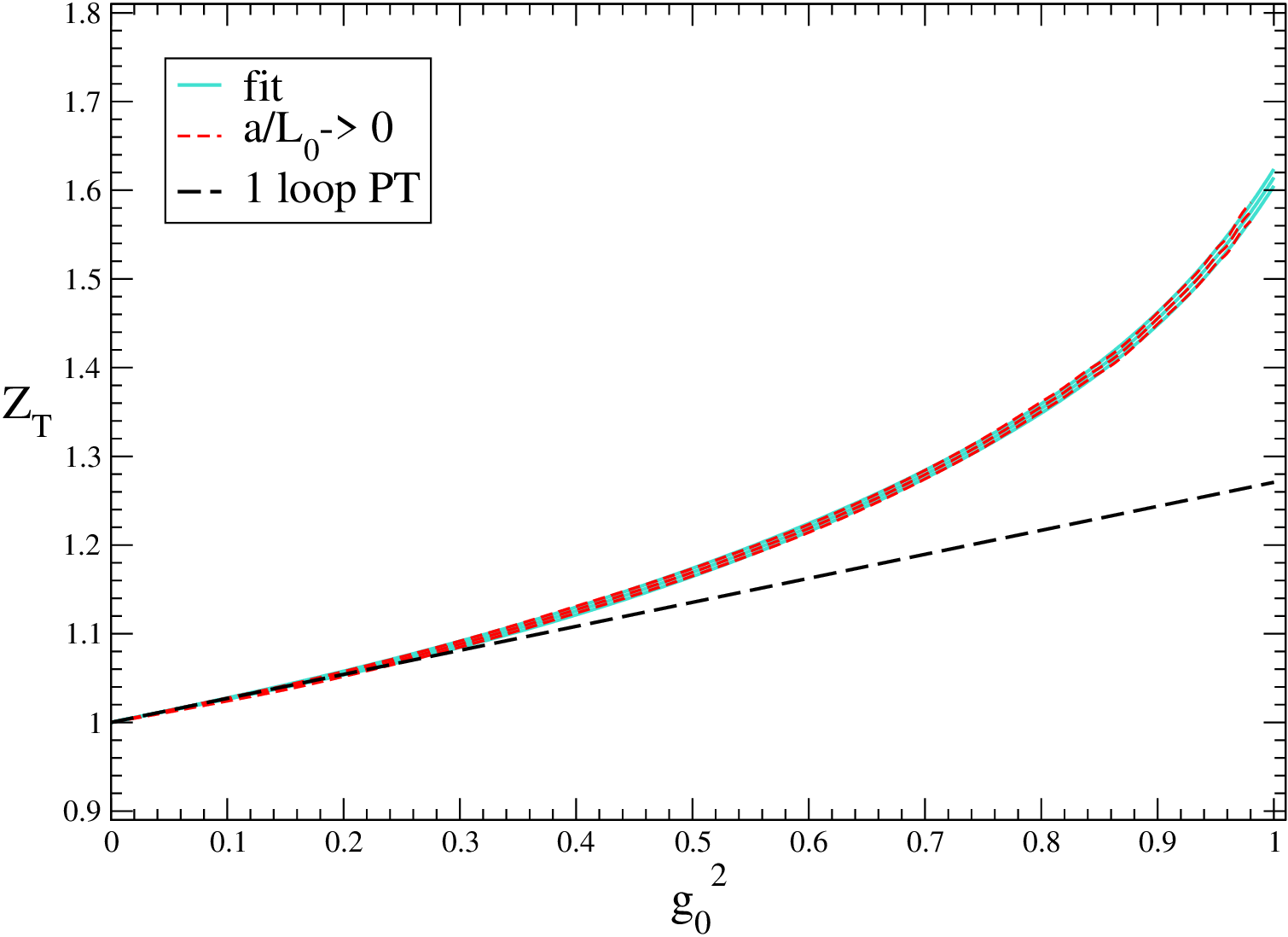}
\caption{
Left: the renormalization factor $Z_{_T}(g_0^2)$ as a function of the 
bare coupling $g_0^2$ for $L_0/a=3$ (blue), $4$ (green) and $5$ (red).
Right: the renormalization factor $Z_{_T}(g_0^2)$ defined in the limit $a/L_0 \rightarrow 0$
together with the fit to the formula in Eq.~(\ref{eq:ZTfinal}) and 
the one-loop analytic result in Eq.~(\ref{eq:PT1loop}).\label{fig:ZT}}
\end{figure*}

\subsection{Determination of $Z_{_T}$\label{sec:Ztmeth}}
The direct determination of $\Delta f/\Delta \xi_k$ in Eq.~(\ref{ZTpractic}) would 
involve the computation of the ratio of two partition functions with different shifts at the
same value of $L_0/a$ and $g_0^2$. Since the relevant phase spaces in the path integral of the 
two systems overlap very poorly, the ratio cannot be estimated in a single Monte Carlo 
simulation. A possible way out is to 
define a series of physical systems with actions which interpolate between the 
two original ones, and then use the Monte Carlo procedure of 
Refs.~\cite{deForcrand:2000fi,DellaMorte:2007zz,Giusti:2010bb}. The calculation, however, 
becomes quickly demanding for large lattices since the numerical cost increases
quadratically with the spatial volume. 

To bypass this problem we can profit from the fact that $\Delta f/\Delta \xi_k$
is a smooth function of $g_0^2$ at fixed values of $L_0/a$ and $L/a$ in the 
range of chosen values. Its derivative with respect to $g_0^2$ can be written as 
\begin{equation}\label{intder}
\frac{d}{d g_0^{2}} \frac{\Delta f}{\Delta \xi_k} 
= \frac{1}{2a L^3 g_0^2}\,
\Big\{\langle S \rangle _{{\vec\xi} - a/L_0 \hat k} -\langle S \rangle _{{\vec\xi} + a/L_0 \hat k}\Big\}\; , 
\end{equation}
where $\langle S \rangle _{{\vec\xi}}$ stands for the expectation value of the action in 
Eq.~(\ref{eq:latS}).  Although the quantities on the r.h.s. of Eq.~(\ref{intder}) have values 
which are close to each other, their difference can be computed at a few permille accuracy 
with a moderate numerical effort. The difference 
$\{\langle S \rangle _{{\vec\xi} - a/L_0 \hat k} -\langle S \rangle _{{\vec\xi} + a/L_0 \hat k}\}$ 
has been computed for ${\vec\xi}=(1,0,0)$ and $L/a=48$
at $63$, $59$ and $48$ values of $g_0^2$  for $L_0/a=3,4$ and $5$ respectively. A sample of 
values is reported in Table~\ref{tab:SmS}, while all of them 
are shown in the left plot of Fig.~\ref{fig:bare_res}. At each value of $L_0/a$ the points are 
interpolated with a cubic spline, and the resulting curve is integrated over $g_0^2$. The 
free-case value is computed analytically by using Eq.~(\ref{eq:freedfdx}), and is added to 
the integral. The systematics induced by the interpolation and the numerical integration of 
the data is negligible with respect to the statistical error. 

To complete the calculation of $Z_{_T}(g_0^2)$, the expectation 
value $\langle T^{[1]}_{0k} \rangle_{\vec\xi}$ is measured in a dedicated
set of simulations. It is computed for ${\vec\xi}=(1,0,0)$ and $L/a=48$
at $66$, $60$ and $38$ values of $g_0^2$ for $L_0/a=3,4$ and $5$ respectively. 
A sample of values is reported in Table~\ref{tab:t0k}, and all of them 
are shown in the right plot of Fig.~\ref{fig:bare_res}. By interpolating the 
results with a cubic spline,
the renormalization constant $Z_{_T}(g_0^2)$ is finally determined by the tree-level 
improved version of Eq.~(\ref{ZTpractic}) given by
\be\label{eq:ZTnum}
Z_{_T}(g_0^2) = - \Big\{ 
\frac{\Delta f}{\Delta \xi_k}\, 
\frac{1}{\langle T^{[1]}_{0k} \rangle_{\vec\xi}} -\mbox{ free case}\Big\}\; .
\ee
The results\footnote{Preliminary results have been reported in 
Ref.~\cite{Giusti:2014tfa}} for $Z_{_T}(g_0^2)$ at $L/a=48$ and 
$L_0/a=3,4$ and $5$ are shown in the left plot of Fig.~\ref{fig:ZT}.
At the larger value of $L_0/a=5$,
discretization effects in $a/L_0$ are within our statistical errors. 
Those due to the finiteness
of $a/L$ have been checked by computing\footnote{At this small volume
we have computed $\Delta f/\Delta \xi_k$ either with the 
method described in this section, or with the Monte Carlo procedure in 
Ref.~\cite{Giusti:2010bb}. The numerical results are in agreement within 
statistical errors.} $Z_{_T}$ at $L/a=16$ and $L_0/a=3$ in the 
full range of $g_0^2$, and at $L_0/a=5$ and $6$ for  $g^2_0>0.85$. The results 
at $L_0/a=3$ for $L/a=16$ and $48$ are statistically compatible,
and their central values differ at most by $0.5\%$ toward the larger values
of $g_0^2$. Since on the lattices with $L_0/a=5$ and $L/a=48$
those effects are expected to be suppressed at least by an 
additional factor of 1/8, we conclude that they are 
well within the statistical errors. We thus quote the values of 
$Z_{_T}(g_0^2)$ at $L_0/a=5$ and $L/a=48$ as our best results in the limit 
$a/L_0 \rightarrow 0$, see right plot of Fig.~\ref{fig:ZT}. Even if 
defined by renormalization conditions which 
differ from ours by discretization effects, our values of $Z_{_T}(g_0^2)$ 
at $g^2_0>0.8$ agree with those in Refs.~\cite{Robaina:2013zmb,Giusti:2014ila} which, 
however, in many cases have a much larger statistical error.
\begin{figure*}[t]
\includegraphics[width=7.0 cm,angle=0]{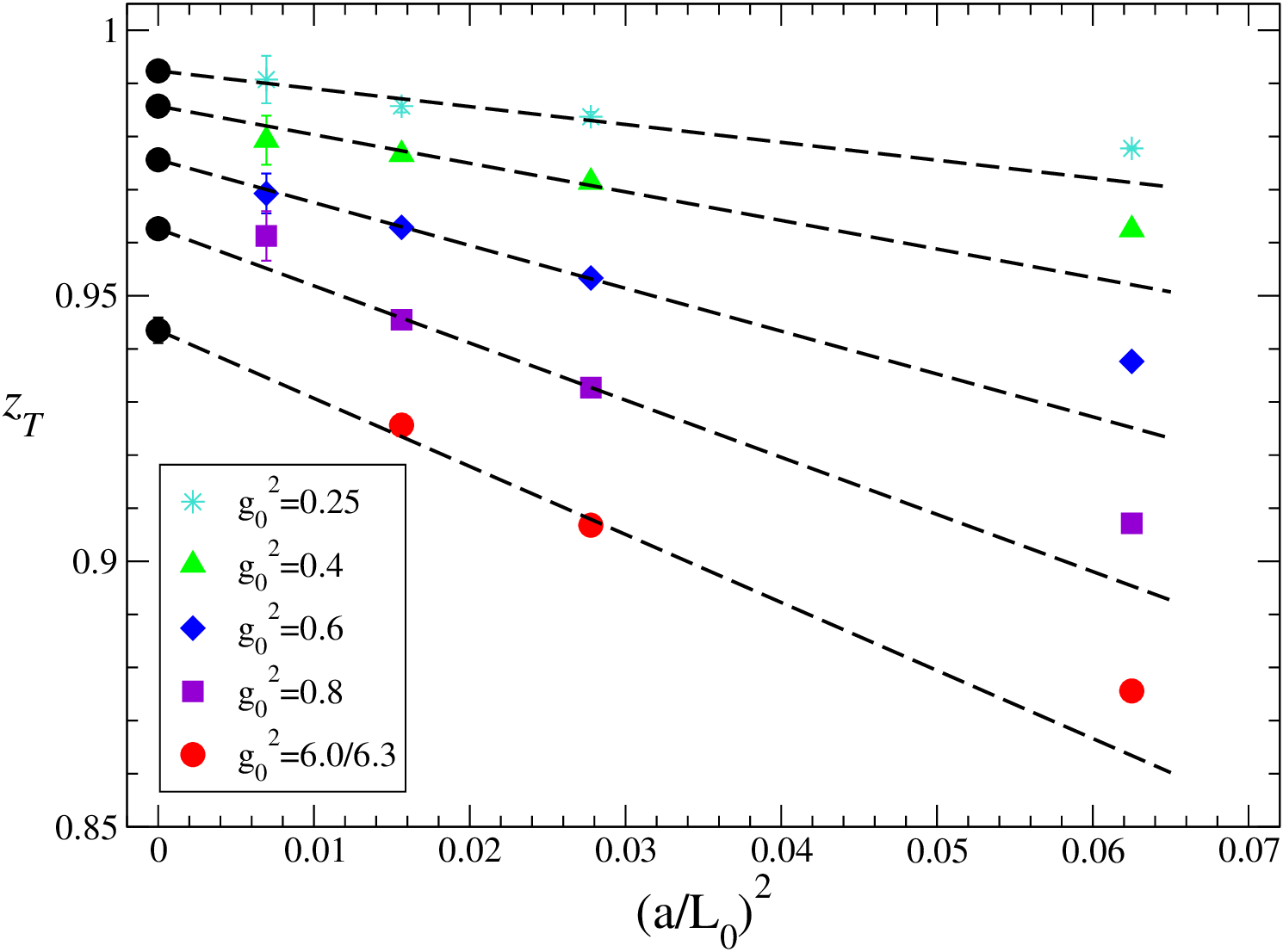}\ \ \ \ \ 
\includegraphics[width=7.0 cm,angle=0]{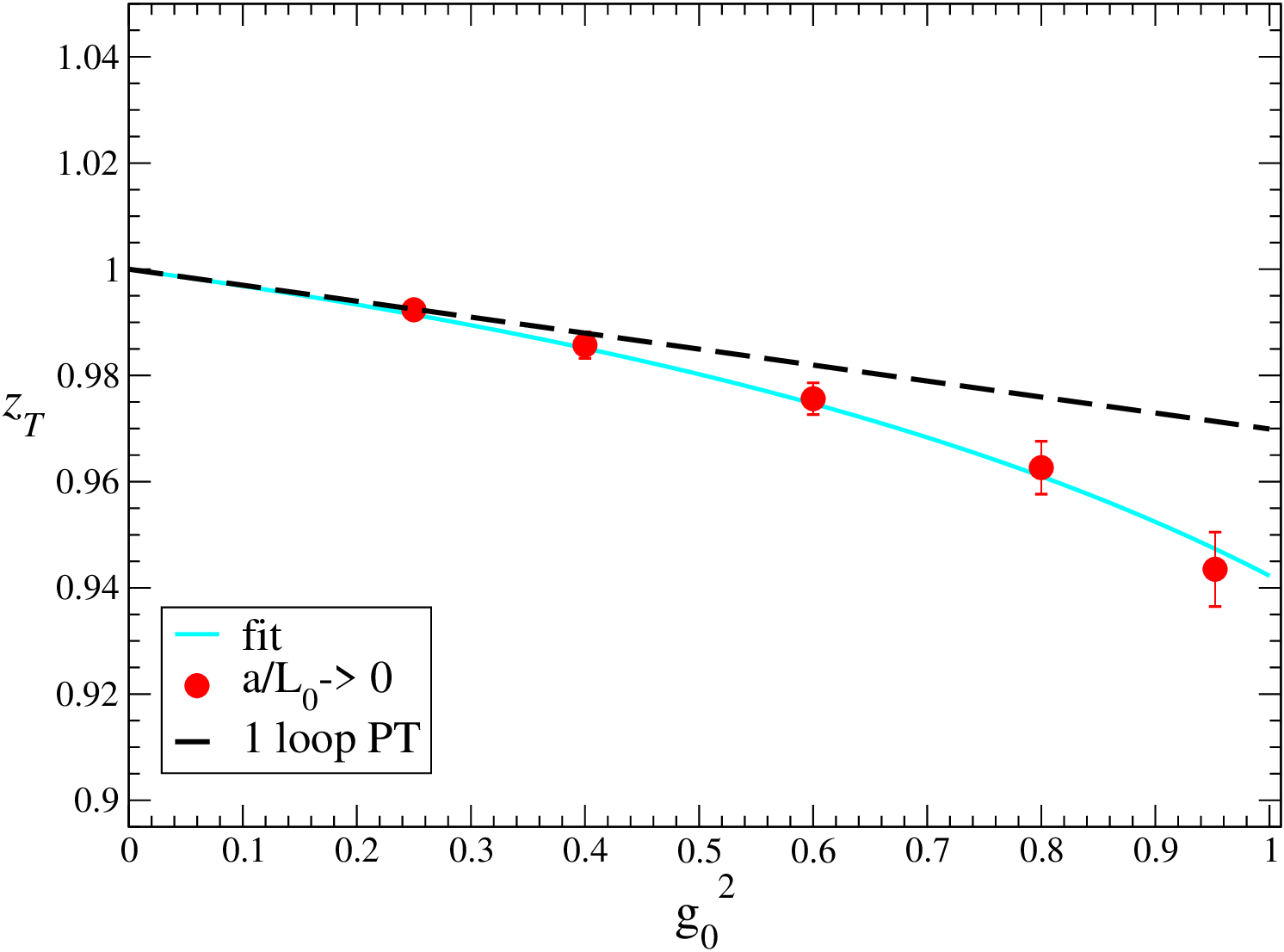}
\caption{Left: the renormalization factor $z_{_T}(g_0^2,a/L_0)$ as a function of 
$(a/L_0)^2$ for the five values of $g_0^2$ indicated in the legend; 
it is also shown the extrapolation to $(a/L_0)=0$ with the fit function in 
Eq.~(\ref{eq:fitzttsmall}). Right: the renormalization 
factor $z_{_T}(g_0^2)$ at $a/L_0=0$ for the 5 values of $g_0^2$ simulated,
together with their fit to the formula in Eq.~(\ref{eq:ztsfinal}) and 
with the one-loop analytic result in Eq.~(\ref{eq:PT1loop}).
\label{fig:Zttsmall}}
\end{figure*}

\subsection{Determination of $z_{_T}$}
The renormalization constant $z_{_T}$ is computed by 
imposing the tree-level improved version of Eq.~(\ref{Zd}) given by
\be\label{eq:ztsmall}
z_{_T}(g_0^2) =  \frac{1-\xi_k^2}{\xi_k}\Big\{ 
\frac {\langle T^{[1]}_{0k} \rangle_{\vec\xi} }{\langle T^{[3]}_{00}\rangle_{\vec\xi} - 
\langle T^{[3]}_{kk} \rangle_{\vec\xi}} - \mbox{ free case}\Big\}\; ,
\ee 
with
\be
\frac{L\, \xi_k}{L_0(1+\xi_k^2)} =  q \in \mathbb{Z}\; . 
\ee 
The latter condition guarantees that the WI remains valid 
at finite volume as it stands~\cite{Giusti:2012yj}.
The expectation values of  
$\langle T^{[1]}_{0k} \rangle_{\vec\xi}$ and of the difference 
$\langle T^{[3]}_{00} \rangle_{\vec\xi} - \langle T^{[3]}_{kk} \rangle_{\vec\xi}$ are 
measured straightforwardly in the same Monte Carlo 
simulation\footnote{It is interesting to notice that the difference
$\langle T^{[3]}_{00} \rangle_{\vec\xi} - \langle T^{[3]}_{kk} \rangle_{\vec\xi}$ requires
roughly 10 times the statistics needed for $\langle T^{[1]}_{0k} \rangle_{\vec\xi}$ 
to meet the same relative statistical error.}. The free case is 
subtracted analytically by using its expression in Appendix~\ref{app:appe}.
In practice we chose ${\vec\xi} = (1/2,0,0)$ and $q=8$ so that 
the ratio of the spatial linear size over the temporal 
one is fixed to be $L/L_0=20$. We simulated 5 values of $g_0^2$ in the range 
$0\leq g_0^2 \leq 1$ with temporal length $L_0/a=4,6,8$ and $12$.
The results for $z_{_T}(g_0^2)$ are given in Table~\ref{tab:ztbare}, and they 
are shown in the left plot of Fig.~\ref{fig:Zttsmall}.  Discretization effects 
turn out to be quite larger than for $Z_{_T}(g_0^2)$ at the smaller values
of $L_0/a$. Our best extrapolation to $a/L_0=0$ is given by the overall fit of 
the data at $L_0/a=6,8$ and $12$ to the function 
\be\label{eq:fitzttsmall}
z_{_T}(g_0^2,a/L_0) = z_{_T}(g_0^2) +  b_1\, g_0^2\, \left(\frac{a}{L_0}\right)^2\; . 
\ee
The quality of the fit is very good, and it leads to the values of $z_{_T}(g_0^2)$ shown 
by the black points in the same plot.
To check for the systematics associated to the extrapolation, we have performed a variety 
of different fits: we have removed the points at $L_0/a=6$ from our best fit, we have fit 
each set of points independently with a quadratic function in $(a/L_0)^2$,  we have 
amended the combined fit function by adding a quadratic 
term in $g^2_0$ to the coefficient of $(a/L_0)^2$, and we have added 
a quadratic term in $(a/L_0)^2$ in Eq.~(\ref{eq:fitzttsmall}). The results of all these fits 
are statistically compatible with those obtained in our best fit to the function in 
Eq.~(\ref{eq:fitzttsmall}) and the selection of data points chosen. We take the maximum spread
of the central values from the various fits as a systematic error due to the extrapolation,
and we add it in quadrature to the statistical one. The final results are shown in the 
right plot of Fig.~\ref{fig:Zttsmall}.

\section{Results and conclusions}
The final results for $Z_{_T}(g_0^2)$ are shown in the right plot of 
Fig.~\ref{fig:ZT}. They are very well represented by the expression
\ba\label{eq:ZTfinal}
Z_{_T}(g_0^2) & = & \frac{1 - 0.4457\, g_0^2}{1 - 0.7165\, g_0^2} - 0.2543\, g_0^4\nonumber\\[0.25cm]
& + & 0.4357\, g_0^6 - 0.5221\, g_0^8
\ea
in the full range $0\leq g_0^2\leq 1$, a function which coincide with the 
expansion in Eq.~(\ref{eq:PT1loop}) to order $g_0^2$. The deviation 
of the curve from the data is smaller than the statistical accuracy, 
see right plot of Fig.~\ref{fig:ZT}. The error to be attached to 
$Z_{_T}(g_0^2)$ computed as in Eq.~(\ref{eq:ZTfinal}) is $0.4\%$ up 
to $g_0^2\leq 0.85$, while it grows linearly from $0.4\%$ to $0.7\%$ 
in the range $0.85\leq g_0^2\leq 1$. Within our statistical errors,
the non-perturbative determination starts to deviate significantly
from the one-loop result at $g_0^2 \sim 0.25$.

Our best results for $z_{_T}(g_0^2)$ are shown in the right plot of 
Fig.~\ref{fig:Zttsmall}. In the full range $0\leq g_0^2\leq 1$,
they are well represented by the expression
\be\label{eq:ztsfinal}
z_{_T}(g_0^2) = \frac{1 - 0.5090\, g_0^2}{1 - 0.4789\, g_0^2}\; , 
\ee
a function which again coincide with the expansion in 
Eq.~(\ref{eq:PT1loop}) to order $g_0^2$. In this case the 
error to be attached to the values in 
Eq.~(\ref{eq:ztsfinal}) grows linearly from  
$0.15\%$ to $0.75\%$ in the interval $0\leq g_0^2\leq 1$. The one loop 
result agrees with the non-perturbative determination up to $g_0^2\sim 0.4$
within our statistical errors.\\

The above results for $Z_{_T}(g_0^2)$ and $z_{_T}(g_0^2)$ clearly show that 
in the range of $g_0^2$ where the Wilson action is 
frequently simulated, one-loop perturbation theory is not adequate for computing   
the renormalization constants of the traceless components of the energy-momentum 
tensor defined as in Eq.~(\ref{eq:lattmunu}). Shifted boundary conditions offer
an extremely powerful tool to compute them, and therefore to 
define non-perturbatively the energy-momentum tensor on the 
lattice. The strategy implemented here can be easily generalized to QCD, 
and to (beyond Standard Model) QCD-like or supersymmetric theories. 

\acknowledgments{
We thank H.~B.~Meyer and D. Robaina for interesting discussions 
in an early stage of this work. The simulations were performed on the BG/Q at 
CINECA (INFN and LISA agreement), on the PC cluster Zefiro of the INFN-CSN4 
in Pisa, and on PC clusters at the Physics Department of the University 
of Milano-Bicocca. We thankfully acknowledge the computer resources and 
technical support provided by these institutions. This work was partially 
supported by the INFN SUMA project.
}

\appendix

\section{SU(3) conventions\label{app:appa}}
The Lie algebra of SU(3) may be identified with the linear space of all
hermitian traceless $3\times 3$ matrices. In the basis 
$T^a$, $a=1 \dots 8$, with 
\be
\Tr[\,T^a] = 0 \; ,\quad T^{a\dagger} = T^a\; ,
\ee
the elements of the algebra are linear combinations
of them with real coefficients. The structure constants $f^{abc}$ in the 
commutator relation  
\be
[T^a,T^b]   =  i f^{abc}T^c 
\ee
are real and totally anti-symmetric in the indices if the normalization
condition
\be
\Tr[\,T^a T^b] = \frac{1}{2}\delta^{ab}
\ee
is imposed. 

\section{Continuum notation\label{app:appb}}
In the Euclidean space-time, the path integral of the SU(3) Yang--Mills 
theory is defined
as 
\be
Z = \int D A\, D \bar c \, D c\; e^{-S}\; , 
\ee
where the measures on gauge and ghost fields are defined as usual. The 
action is defined as
\be\label{eq:SQCD}
S=\int d^4 x\, {\cal L}(x)\; , \qquad {\cal L} = {\cal L}^G + {\cal L}^{GF}\; , 
\ee 
with 
\ba
{\cal L}^G & = & \frac{1}{2 g_0^2} \Tr\Big[F_{\mu\nu}\, F_{\mu\nu} \Big]\; , \\ 
{\cal L}^{GF} & = & \frac{\lambda_0}{g^2_0} \Tr\Big[\partial_\mu A_\mu\, \partial_\nu A_\nu\Big]
+ \frac{2}{g_0^2} \Tr\Big[\partial_\mu \bar c\, {\cal D}_\mu c \Big]\nonumber
\ea
where $g_0$ is the bare coupling constant, $\lambda_0$ is the gauge-fixing 
parameter, the trace is over the color index and  
\ba\label{eq:basics}
F_{\mu\nu}  & = & \partial_\mu A_\nu - 
\partial_\nu A_\mu -i\, [A_\mu,A_\nu]\; , \\ 
{\cal D}_\mu\, c & = & \partial_\mu\, c - i\, [A_\mu,c]\; ,
\qquad A_\mu=A_\mu^a\, T^a \; . \nonumber
\ea
The ghost fields $c$ and $\bar c$ are in the adjoint representation of the 
SU(3) group, i.e. $c =c^a  T^a$ and analogously for $\bar c$. 

\subsection{BRST transformations}
The action (\ref{eq:SQCD}) is invariant under the BRST transformations 
defined as~\cite{Becchi:1974xu,Becchi:1974md,Tyutin:1975qk}
\ba\label{eq:BRST}
\delta A_{\mu} & = & \theta\, D_\mu c\nonumber\\[0.125cm]
\delta \bar c & = & \lambda_0\, \theta\, (\partial_\mu A_\mu)\\[0.125cm]
\delta c & = & i \theta\, c^2\nonumber
\ea
where $\theta$ is an infinitesimal Grassmann constant. They are nilpotent 
up to the equations of motion of the ghost field $c$. In fact if we define 
\be
\delta \phi = \phi' - \phi = \theta \Delta \phi\; , 
\ee
where $\phi$ is one of the fundamental fields which transforms as in 
Eqs. (\ref{eq:BRST}), it is easy to prove that\footnote{To this aim it is 
useful to notice that $\delta (D_\mu c)=0$.} 
\be\label{eq:nilFO}
\Delta^2 A_\mu=0\; , \quad \Delta^2 c = 0\; , 
\ee 
while 
\be\label{eq:nilcbar}
\Delta^2 \bar c = \lambda_0\, \partial_\mu D_\mu\, c\; . 
\ee
By using Eqs.~(\ref{eq:nilFO}) and (\ref{eq:nilcbar}) it is 
easy to show that the BRST transformations are nilpotent, up to the 
equation of motion of $c$, when acting on any product of fundamental  
fields at arbitrary space-time points and thus on any functional
of them.

The gauge-invariant part of the Yang--Mills action (\ref{eq:SQCD}) is 
BRST-invariant because the BRST correspond to infinitesimal gauge 
transformations with parameter $\theta c(x)$. The gauge-fixing part of the action 
turns out to be BRST-invariant too. It can also be written as a BRST rotation 
of a functional plus a term which, after integrating by parts, is proportional 
to the equation of motion of $c$ and serves to cancel the term coming from 
Eq.~(\ref{eq:nilcbar}). 

\subsection{Equations of motion}
 The equations of motion for the gauge field are given by 
\begin{widetext}
\ba\label{eq:EOMG}
\hspace{-1.25cm}\Big\langle \Big\{\frac{1}{g_0^2}\,[{\cal D}_\alpha F_{\alpha\mu}]^a + \frac{\lambda_0}{g_0^2}\, 
\partial_\mu\partial_\alpha A_\alpha^a + \frac{1}{g_0^2}\, f^{abc}\, 
(\partial_\mu \bar c^b)\, c^c \Big\}(x)\, O \Big\rangle = - 
\Big\langle \frac{\delta O}{\delta A_\mu^a(x)} \Big\rangle\; ,
\ea
\end{widetext}
where $O$ represents a generic string of fields, and the covariant 
derivative for the adjoint representation is defined as in 
Eq.~(\ref{eq:basics}), i.e.  
\be
{\cal D}_\mu\, F_{\mu\nu} = \partial_\mu\, F_{\mu\nu} - i\, [A_\mu,F_{\mu\nu}]\; .
\ee
The equations of motion for the ghosts are 
\ba
\Big\langle \frac{1}{g_0^2} \partial_\mu [{\cal D}_\mu\, c]^a(x) \, O\Big\rangle & = &  
- \Big\langle \frac{\delta}{\delta \bar c^a(x)}\, O\Big\rangle\; , \\
\Big\langle \frac{1}{g_0^2} [{\cal D}_\mu\partial_\mu\bar c]^a (x)\, O\Big\rangle & = & 
\Big\langle \frac{\delta}{\delta c^a(x)}\, O\Big\rangle\; .\nonumber
\ea

\section{Energy-momentum tensor in the continuum\label{app:appc}}
The continuum theory is invariant under the group of space-time translations
\be\label{eq:transl}
x_\mu' = x_\mu - \varepsilon_\mu\; , \qquad \phi'(x_\mu') = \phi(x_\mu)\, ,
\ee
where $\phi$ indicates generically one of the fields $A_\mu, c, \bar c$.
The associated WIs can be derived in the usual way by studying the variation of the 
functional integral under local transformations parameterized by $\varepsilon_\mu(x)$
\ba\label{eq:trdelta}
\delta A_\mu(x) & = & \varepsilon_\rho(x)\, \partial_\rho A_\mu(x)\, ,\nonumber\\[0.25cm]
\delta c(x) & = & \varepsilon_\rho(x)\, \partial_\rho c(x)\; , \quad 
\delta \bar c(x) =\bar c(x)\overleftarrow \partial_\rho\, \varepsilon_\rho(x)\; .
\ea
The resulting integrated WIs are 
\be\label{eq:WIg}
\int d^4 z\, \varepsilon_\nu(z)\, \langle \partial_\mu T^c_{\mu\nu}(z)\, 
O \rangle = - \left\langle \delta O \right\rangle\, ,
\ee
where $\delta O$ is the variation of the string of fields $O$ under the transformation 
(\ref{eq:trdelta}). The canonical energy-momentum tensor of the theory can be 
written as 
\be\label{eq:Tc}
T^c_{\mu\nu} = T_{\mu\nu}^{G,c} + T_{\mu\nu}^{GF,c}\; ,
\ee
where 
\ba\displaystyle
T_{\mu\nu}^{G,c} & = & \frac{2}{g_0^2}\, \Tr\Big[F_{\mu\alpha} \partial_\nu A_\alpha\Big]
- \delta_{\mu\nu}\, {\cal L}^G\; ,\\[0.125cm]
T_{\mu\nu}^{GF,c} & = & \frac{2\lambda_0}{g_0^2}\Tr\Big[\partial_\alpha A_\alpha 
\partial_\nu A_\mu\Big]\\ 
& + & \frac{2}{g_0^2}\Tr\Big[(\partial_\mu\bar c)(\partial_\nu c)
+ (\partial_\nu \bar c)(D_\mu c)\Big] - \delta_{\mu\nu}\,{\cal L}^{FG}\; .\nonumber  
\ea
For $\varepsilon_\nu(z) = \epsilon_\nu \delta^{(4)}(z-x)$,  Eq.~(\ref{eq:WIg}) gives
\be\label{eq:WIstd1}
\epsilon_\nu\, \langle \partial_\mu T^c_{\mu\nu}(x)\, O \rangle = -  
\left\langle \delta_x O \right\rangle\; ,
\ee
and when all operators of the string $O$ are localized far away from $x$, the 
classical conservation identities  
\be\label{eq:WIstd2}
\langle \partial_\mu T^c_{\mu\nu}(x)\, O \rangle = 0\;   
\ee
are recovered. The canonical energy-momentum tensor is neither symmetric nor 
gauge invariant. To make it both symmetric and gauge invariant one applies 
the Belinfante procedure, and use the equation of motion. The resulting  
tensor satisfies the on-shell WIs in Eq.~(\ref{eq:WIstd2}), and it gives the 
same conserved charges of the canonical tensor when inserted in on-shell 
correlation functions. The ambiguity left by the use of the equations of motion 
allows one to define the energy-momentum tensor as the one derived by exploiting 
the re-parameterization transformations of the theory coupled to an 
external gravitational field \cite{Callan:1970ze,Adler:1976zt,Collins:1976yq}. 
All definitions related by terms which vanish 
by the equation of motion are equivalent provided the corresponding contact 
terms are taken into account in the WIs. 
The symmetric energy-momentum tensor is defined as 
\be\label{eq:Theta}
T^B_{\mu\nu} = T_{\mu\nu}^{G,B} + T_{\mu\nu}^{GF,B}\, ,
\ee
where 
\begin{widetext}
\ba\displaystyle
T_{\mu\nu}^{G,B} & = & \frac{2}{g^2_0}\Tr\Big[F_{\mu\alpha}F_{\nu\alpha}\Big] 
- \delta_{\mu\nu}\, {\cal L}^G\, ,\\[0.25cm]
T_{\mu\nu}^{GF,B} & = & \frac{2\lambda_0}{g_0^2}\Tr\left[-A_\mu \partial_\nu\partial_\alpha 
A_\alpha - A_\nu \partial_\mu \partial_\alpha A_\alpha + \delta_{\mu\nu} 
\Big(\frac{1}{2} \partial_\alpha A_\alpha \partial_\beta A_\beta  + 
A_\alpha \partial_\alpha \partial_\beta A_\beta\Big)\right] + \nonumber\\[0.25cm]
& & \frac{2}{g_0^2}\Tr\Big[\partial_\mu\bar c\, D_\nu c
+ \partial_\nu \bar c\, D_\mu c\Big] - 
\delta_{\mu\nu}\, \frac{2}{g_0^2} \Tr\Big[\partial_\alpha \bar c\, D_\alpha c \Big]
\; . \nonumber  
\ea
By comparing Eqs.~(\ref{eq:Tc}) and (\ref{eq:Theta}), it is quite easy to 
show that 
\ba
\partial_\mu T^c_{\mu\nu} =  
\partial_\mu T^{B}_{\mu\nu} & + &  
\partial_\mu\left\{A_\nu^a\, \Big[\frac{1}{g_0^2}[D_\alpha F_{\alpha\mu}]^a 
+ \frac{\lambda_0}{g_0^2}\, \partial_\mu\partial_\alpha A^a_\alpha
+ \frac{1}{g_0^2} f^{abc} (\partial_\mu \bar c^b)\, c^c\Big]\right\}\, ,
\ea
\end{widetext}
i.e. the two four-divergences differ by terms which are proportional to 
the equations of motion. If we insert last equation in the WIs 
(\ref{eq:WIstd1}) and we use the 
the equations of motion (\ref{eq:EOMG}) we arrive to 
\be\label{eq:WIbel}
\epsilon_\nu \langle \partial_\mu  T^B_{\mu\nu}(x)\, O \rangle\!
=\!- \left\langle \delta_x O  \right\rangle\! 
+\!\epsilon_\nu \partial_\mu \Big\langle A_\nu^a(x) \frac{\delta O}{\delta A^a_\mu(x)}
\Big\rangle\, .\! 
\ee
It is also useful to notice that 
\be
T_{\mu\nu}^{GF,B} = \Delta\, \Xi_{\mu\nu} + \delta_{\mu\nu}\, \frac{1}{g_0^2}\, 
\Tr\Big[\bar c\, \partial_\alpha D_\alpha\, c\Big]\; , 
\ee
where $\Delta$ is the BRST variation defined in appendix~\ref{app:appb} and 
\ba
\Xi_{\mu\nu} & = & \frac{2}{g_0^2}\Tr\Big[-A_\mu\partial_\nu\bar c - A_\nu\partial_\mu \bar c +\\
& &  \delta_{\mu\nu}\Big(\frac{1}{2}(\partial_\alpha A_\alpha)\,\bar c + A_\alpha \partial_\alpha \bar c 
\Big) \Big]\; . \nonumber
\ea
When the interpolating operator $O$ is gauge-invariant, 
it is thus appropriate to define a gauge-invariant energy-momentum
tensor 
\be\label{eq:GITmunu}
T_{\mu\nu} = T_{\mu\nu}^{G,B} = \frac{1}{g^2_0}\Big\{F^a_{\mu\alpha}F^a_{\nu\alpha} 
- \frac{1}{4}\delta_{\mu\nu}\, F^a_{\alpha\beta}F^a_{\alpha\beta}\Big\}\\[0.25cm]
\ee
which satisfies 
\be\label{eq:corGI}
\langle \partial_\mu  T_{\mu\nu}(x)\, O \rangle = 
\langle \partial_\mu  T^B_{\mu\nu}(x)\, O \rangle\; ,  
\ee
where the term proportional to the equation of motion of the ghosts is null because
a gauge-invariant operator is independent of the $\bar c$ field. The WIs 
(\ref{eq:WIbel}) applies as well to $T_{\mu\nu}(x)$ without any modification.
It is worth nothing that the gauge-invariant energy-momentum tensor generates 
the very same charges in on-shell correlation functions as all the other 
definitions in this appendix.

\section{Renormalization of the action density in dimensional regularization \label{app:appd}}
In this appendix we report the essential formulas in dimensional regularization which are 
needed in the paper, for a recent review see Ref.~\cite{Weisz:2010nr} and reference therein. 
In dimensional regularization one replaces $\int d^4 x \rightarrow \int d^D x$, and 
renormalizes the coupling constant as 
\be
g_0^2 = \mu^{2\epsilon}\, g^2 Z^{-1}_g\, ,
\ee
where $D=4-2\epsilon$. The $\beta$-function is 
\ba
\tilde\beta(\epsilon, g)  & = & \mu \frac{\partial g}{\partial \mu} = 
-\epsilon g \left\{1 - \frac{g}{2}\frac{\partial}{\partial g} \ln{Z_g} \right\}^{-1}\nonumber\\
 & = & -\epsilon g + \beta(g)\; ,
\ea
where
\be
\beta(g) = -g^3 \sum_{k=0}^{\infty} b_k g^{2k}\; , 
\ee
and
\be\label{eq:bo1}
b_0 = \frac{1}{(4\pi)^2} \frac{11}{3} N_c\; \qquad 
b_1 = \frac{1}{(4\pi)^4} \frac{34}{3} N^2_c\; 
\ee
with the number of colours being $N_c=3$ in our case.

In presence of shifted boundary conditions it holds 
\be
\frac{\partial}{\partial g} \langle T_{0k} \rangle_{\vec\xi} = 
\frac{1}{L_0} \frac{\partial g_0}{\partial g} \frac{1}{2g_0^3}
\frac{\partial}{\partial\xi_k} \langle F^a_{\alpha\beta} F^a_{\alpha\beta} \rangle_{\vec\xi}\; , 
\ee
which can be written as ($x_0\neq 0$)

\begin{widetext}
\be\label{eq:ddd}
\frac{\partial}{\partial g} \langle T_{0k} \rangle_{\vec\xi} = \frac{1}{2}
\frac{\partial g_0}{\partial g} \frac{g^3}{g_0^3} Z_{E} \left[\frac{1}{g^3} 
\langle {\overline T}_{0k}(x_0)\; \{F^a_{\alpha\beta} F^a_{\alpha\beta}(0)\}^{\rm R} \rangle_{\vec\xi\,,c}\right]\; , 
\ee
\end{widetext}
where $Z_E$ and the renormalized density are defined in Eq.~(\ref{eq:ZE}).
The expectation values of the renormalized operators in Eq.~(\ref{eq:ddd}) are finite and 
expandable in powers of $g$. By following Ref.~\cite{Luscher:2013lgggg}, see 
also Ref.~\cite{Suzuki:2013gza}, the ratio
\be
\mu^{2\epsilon} \frac{\partial g_0}{\partial g} \frac{g^3}{g_0^3} Z_{E}
=-\epsilon g \frac{Z_{E} Z_g}{\tilde \beta(\epsilon,g)}
\ee
must then have a series in $g$ with no poles in $\epsilon$. In dimensional regularization 
the coefficients of the poles in $\epsilon$ in the renormalization constants 
start at $O(g^2)$, and therefore
\be
\epsilon Z_E Z_g = \epsilon + R\; , \qquad R = \sum_{k=1}^{\infty} r_k g^{2k}\; , 
\ee
which implies 
\be
-\epsilon g \frac{Z_{E} Z_g}{\tilde\beta(\epsilon,g)} = 
\frac{g (\epsilon + R)}{\epsilon g -\beta(g)}\; . 
\ee
If we expand in $g$ the denominator we obtain
\be
\frac{g (\epsilon + R)}{\epsilon g -\beta(g)} = \left(1+ \frac{R g}{\beta}\right)\sum_{k=0}^{\infty} 
\left(\frac{\beta}{\epsilon g}\right)^k - \frac{R g}{\beta}\; . 
\ee
Since this quantity cannot have poles in $\epsilon$
\be
R=-\frac{\beta}{g}\; , 
\ee
and therefore 
\be\label{eq:ZEr}
\epsilon Z_{E} Z_g = \epsilon - \frac{\beta}{g}\; . 
\ee
The renormalization constant of the energy-density operator
is unambiguously fixed from the one of the coupling.

\section{Lattice free theory with shifted boundary conditions\label{app:appe}}
In this appendix 
we report the results for the expectation values 
of $\langle T^{[1]}_{0k} \rangle_{\vec\xi}$, 
$\langle T^{[3]}_{00}\rangle_{\vec\xi} - \langle T^{[3]}_{kk} \rangle_{\vec\xi}$ 
(no sum over k), $\langle T^{[2]}_{\mu\mu} \rangle_{\vec\xi}$, and $\Delta f/\Delta \xi_k$ 
in the free theory on the lattice\footnote{The lattice spacing is set to $a=1$ in this 
appendix.}. In the infinite volume limit the 
expectation value of the momentum density is given by~\cite{Giusti:2012yj}
\begin{widetext}
\be
\langle T^{[1]}_{0k} \rangle_{\vec\xi} = \frac{8}{L_0} \sum_{\ell=0}^{L_0-1} \int_{BZ} \frac{d^3\vec p}{(2\pi)^3}
\; \frac{\sin(p_0)\sin(p_k)\sum_{\alpha\neq 0,k} \cos^2(p_\alpha/2)}
{4\sin^2(\frac{p_0}{2}) 
+ \omega^2_{\vec p}}\; , 
\ee
where 
\be\label{eq:subs}
\phi_{\vec p} = \vec p\cdot\vec \xi\; , \qquad
\omega^2_{\vec p}=4 \sum_{k=1}^3 \sin^2(\frac{p_k}{2})\; , \qquad
p_0 = \frac{2\pi\ell}{L_0} - \phi_{\vec p}\, .
\ee 
If we notice that~\cite{Giusti:2012yj}
\ba
\Sigma(\phi,\omega,x_0)& = &\frac{1}{L_0}\sum_{\ell=0}^{L_0-1} 
\frac{e^{ix_0(2\pi \ell/L_0 -  \phi)}}
{4\sin^2(\frac{\pi\ell}{L_0} - \frac{\phi}{2}) + \omega^2}\nonumber\\[0.25cm] 
& = & 
\frac{1}{2\sinh\hat\omega}\left[  
\frac{e^{\hat\omega x_0}}{e^{iL_0\phi+L_0\hat\omega}-1} - 
\frac{e^{-\hat\omega x_0}}{e^{iL_0\phi-L_0\hat\omega}-1} \right]
\; ,
\ea
where $\omega = 2\sinh(\hat\omega/2)$, and that real and imaginary 
parts of $\Sigma$ read
\ba\label{eq:reSig}
\re\Sigma(\phi,\omega,x_0)&=& 
\frac{\sinh(L_0\hat\omega/2)\cosh[\hat\omega(L_0/2-x_0)] - \sin^2(L_0\phi/2) \sinh(\hat\omega x_0)}
{\sinh(\hat\omega)\; \left(\cosh(L_0\hat\omega) - 
\cos(L_0\phi)\right)}\;,\\[0.25cm]
\im\Sigma(\phi,\omega,x_0)&=& \frac{-\sin(L_0\phi)\,\sinh(\hat\omega x_0)}
{2\sinh(\hat\omega)\,\left(\cosh(L_0\hat\omega) - \cos(L_0\phi)\right)}
\; ,
\label{eq:imSig}
\ea
we arrive to
\be
\langle T^{[1]}_{0k} \rangle_{\vec\xi} = 8
\int_{BZ} \frac{d^3\vec p}{(2\pi)^3}  
\sin(p_k)\; \im\Sigma(\phi_{\vec p},\omega_{\vec p},1) \sum_{\alpha\neq 0,k} \cos^2(\frac{p_\alpha}{2})\; .
\ee
Analogously, for the traceless diagonal component of the energy-momentum tensor
we obtain 
\ba
& &\hspace{2.0cm} \langle T^{[3]}_{00}\rangle_{\vec\xi} - \langle T^{[3]}_{kk} \rangle_{\vec\xi} =  
\frac{4}{L_0} \sum_{\ell=0}^{L_0-1} \int_{BZ} \frac{d^3\vec p}{(2\pi)^3}\; 
\frac{1}{4\sin^2(\frac{p_0}{2}) 
+ \omega^2_{\vec p}}\times\\[0.25cm]
& &\hspace{-0.5cm} \Big\{ \Big[\cos(p_0) - \cos(p_k)\Big]
\sum_{\alpha\neq 0,k} \sin^2(p_\alpha) -
\Big[\cos(2 p_0) - \cos(2 p_k)\Big] 
\sum_{\alpha\neq 0,k} \cos^2(\frac{p_\alpha}{2})
\Big\}\, , \nonumber
\ea
which by summing over $l$ gives
\ba
& &\hspace{-0.5cm}\langle T^{[3]}_{00}\rangle_{\vec\xi} - \langle T^{[3]}_{kk} \rangle_{\vec\xi} 
=  4\!\!\int_{BZ} \frac{d^3\vec p}{(2\pi)^3}
\Big\{\re\Sigma(\phi_{\vec p},\omega_{\vec p},1)\!\!\!\sum_{\alpha\neq 0,k} \sin^2(p_\alpha)
    - \re\Sigma(\phi_{\vec p},\omega_{\vec p},2)\!\!\!\sum_{\alpha\neq 0,k} \cos^2(\frac{p_\alpha}{2})\nonumber\\[0.25cm]
& & \hspace{2.2cm}
+ \re\Sigma(\phi_{\vec p},\omega_{\vec p},0)\Big[\cos(2 p_k)\!\!\!\sum_{\alpha\neq 0,k} \cos^2(\frac{p_\alpha}{2}) 
- \cos(p_k)\!\!\!\sum_{\alpha\neq 0,k} \sin^2(p_\alpha) 
\Big]
\Big\} \; .  
\ea
For the trace part we obtain
\be
\langle T^{[2]}_{\mu\mu} \rangle_{\vec\xi} =  -
\frac{16}{L_0} \sum_{\ell=0}^{L_0-1} \int_{BZ} \frac{d^3\vec p}{(2\pi)^3}\; 
\frac{\sum_{\alpha, \beta\neq\alpha} \cos^2(\frac{p_\alpha}{2}) \sin^2(p_\beta)}
{4\sin^2(\frac{p_0}{2}) + \omega^2_{\vec p}} \; , 
\ee
which by summing over $l$ gives
\ba
& & \langle T^{[2]}_{\mu\mu} \rangle_{\vec\xi} =
8 \int_{BZ} \frac{d^3\vec p}{(2\pi)^3}\;
\Big\{\re\Sigma(\phi_{\vec p},\omega_{\vec p},2) \sum_{k=1}^3\cos^2(\frac{p_k}{2})-
      \re\Sigma(\phi_{\vec p},\omega_{\vec p},1) \sum_{k=1}^3 \sin^2(p_k)\nonumber\\[0.25cm]
& & \hspace{1.125cm}
- \re\Sigma(\phi_{\vec p},\omega_{\vec p},0) \sum_{k=1}^3\Big[\sin^2(p_k) + \cos^2(\frac{p_k}{2}) 
+2 \cos^2(\frac{p_k}{2}) \sum_{q\neq k} \sin^2(p_q) \Big] \Big\} \; .
\ea 
In the free theory the discrete derivative of the free energy in
Eq.~(\ref{eq:discrder}) is given by
\be
\frac{\Delta f}{\Delta \xi_k} = 4 \sum_{\ell=0}^{L_0-1} \int_{BZ} 
\frac{d^3\vec p}{(2\pi)^3}\;
\ln\left[\frac{\omega^2_{\vec p} + 4 \sin^2((p_0 - p_k/L_0)/2)}
              {\omega^2_{\vec p} + 4 \sin^2((p_0 + p_k/L_0)/2)}
\right]\; , 
\ee
which by summing over $l$ gives
\be\label{eq:freedfdx}
\frac{\Delta f}{\Delta \xi_k} = 2 \int_{BZ} 
\frac{d^3\vec p}{(2\pi)^3}\;
\ln\left[\frac{\cosh(L_0\, \hat\omega_{\vec p}) - \cos(L_0\, \phi_{\vec p} + p_k) }
             {\cosh(L_0\, \hat\omega_{\vec p}) - \cos(L_0\, \phi_{\vec p} -  p_k)}\right]\; . 
\ee
All previous equations remain valid in finite volume if one makes the 
substitution
\be
\int_{BZ} \frac{d^3\vec p}{(2\pi)^3}\rightarrow \frac{1}{V}\sum_{\vec p}\, ,
\ee
and defines a prescription for the zero mode.
\end{widetext}

\section{Numerical results \label{app:appf}}
%\TABLE{
\begin{table}[htb]
\begin{center}
\begin{tabular}{| l | l | l | l |}
\hline
$\beta$ &  \multicolumn{3}{|c|}{$\displaystyle \frac{1}{18} \sum_{\mu,\nu<\mu}\Big[ 
\langle {\rm Re}\Tr\, U_{\mu\nu}\rangle _{{\vec\xi} + a/L_0 \hat k} -
\langle {\rm Re}\Tr\, U_{\mu\nu}\rangle _{{\vec\xi} - a/L_0 \hat k}\Big]$} \\
\hline
&   $L_0/a=3$ &   $L_0/a=4$ &   $L_0/a=5$ \\
\hline
 6.0  &  5.489(10)  $\times 10^{-4}$  &  3.028(7)   $\times 10^{-4}$   &   4.2564(33) $\times 10^{-4}$ \\
 6.03 &  4.886(10)  $\times 10^{-4}$  &  2.443(6)   $\times 10^{-4}$   &   3.8484(38) $\times 10^{-4}$ \\
6.125 &  3.601(14)  $\times 10^{-4}$  &  1.491(7)   $\times 10^{-4}$   &   1.0016(28) $\times 10^{-4}$ \\
 6.5  &  1.576(11)  $\times 10^{-4}$  &  5.160(37)  $\times 10^{-5}$   &   2.339(20)  $\times 10^{-5}$ \\
 7.0  &  7.65(8)    $\times 10^{-5}$  &  2.232(33)  $\times 10^{-5}$   &   8.88(11)   $\times 10^{-6}$ \\
 8.0  &  2.96(5)    $\times 10^{-5}$  &  7.08(25)   $\times 10^{-6}$   &   2.43(9)    $\times 10^{-6}$ \\
 9.0  &  1.604(38)  $\times 10^{-5}$  &  3.60(17)   $\times 10^{-6}$   &   1.09(11)   $\times 10^{-6}$ \\
10.0  &  1.041(23)  $\times 10^{-5}$  &  2.07(8)    $\times 10^{-6}$   &   6.6(9)     $\times 10^{-7}$ \\
11.0  &  7.77(22)   $\times 10^{-6}$  &  1.49(8)    $\times 10^{-6}$   &   3.8(6)     $\times 10^{-7}$ \\
12.0  &  5.85(25)   $\times 10^{-6}$  &  1.06(6)    $\times 10^{-6}$   &   3.2(5)     $\times 10^{-7}$ \\
13.5  &  4.29(35)   $\times 10^{-6}$  &  7.9(8)     $\times 10^{-7}$   &   2.0(6)     $\times 10^{-7}$ \\
17.0  &  2.39(14)   $\times 10^{-6}$  &  3.8(5)     $\times 10^{-7}$   &   1.3(4)     $\times 10^{-7}$ \\
20.0  &  1.52(12)   $\times 10^{-6}$  &  2.8(4)    $\times 10^{-7}$   &   8.2(28)    $\times 10^{-8}$ \\
24.0  &  1.14(8)    $\times 10^{-6}$  &  2.16(27)   $\times 10^{-7}$   &   4.6(24)    $\times 10^{-8}$ \\
30.0  &  7.5(6)     $\times 10^{-7}$  &  1.48(26)   $\times 10^{-7}$   &   5.5(14)    $\times 10^{-8}$ \\
50.0  &  2.99(23)   $\times 10^{-7}$  &  6.6(8)     $\times 10^{-8}$   &   2.3(7)     $\times 10^{-8}$ \\
80.0  &  1.25(15)   $\times 10^{-7}$  &  2.09(39)   $\times 10^{-8}$   &   0.97(37)   $\times 10^{-8}$ \\
 \hline
\end{tabular}
\end{center}
\caption{Values of the difference of the average plaquettes measured at bare coupling $\beta=6/g_0^2$ on lattices of
size $48^3\times L_0/a$. \label{tab:SmS}}
\end{table}
%}
For a representative sample of values of 
$g_0^2$ that we have simulated we collect the results for
the difference of the average plaquettes entering Eq.~(\ref{intder})
in Table \ref{tab:SmS}, and the values of 
$\langle T^{[1]}_{0k} \rangle_{\vec\xi}$ at $\vec\xi=(1,0,0)$ in Table~\ref{tab:t0k}. 
The values of $\langle T^{[1]}_{0k} \rangle_{\vec\xi}$
and $\langle T^{[3]}_{00}\rangle_{\vec\xi} - \langle T^{[3]}_{kk} \rangle_{\vec\xi}$
for ${\vec\xi} = (1/2,0,0)$ are given in Table~\ref{tab:ztbare}.

%\TABLE{
\begin{table}[htb]
\begin{center}
\begin{tabular}{| l | l | l | l |}
\hline
$\beta$ &  \multicolumn{3}{|c|}{$\langle T^{[1]}_{0k} \rangle_{\vec\xi} $} \\
\hline
&   $L_0/a=3$ &   $L_0/a=4$ &   $L_0/a=5$ \\
&   {\small $(\times 10^{-3})$} & {\small $(\times 10^{-3})$} & {\small
$(\times 10^{-4})$} \\
\hline
6.0     &   -5.2735(27)   &   -1.3772(13)  &   -2.826(9)\\
6.03    &   -5.3921(29)   &   -1.4447(11)  &   -4.047(6)\\
6.125   &   -5.6976(29)   &   -1.6064(13)  &   -5.568(5)\\
6.3     &   -6.1359(37)   &   -1.7977(12)  &   -6.797(6)\\
6.5     &   -6.5124(28)   &   -1.9495(12)  &   -7.610(7)\\
7.0     &   -7.1554(29)   &   -2.1899(20)  &   -8.786(7)\\
8.0     &   -7.9077(37)   &   -2.4488(20)  &   -9.916(8)\\
9.0     &   -8.3673(21)   &   -2.5947(30)  &   -10.550(7)\\
10.0    &   -8.6896(14)   &   -2.7002(32)  &   -10.981(8)\\
11.0    &   -8.9385(16)   &   -2.7780(31)  &   -11.288(6)\\
12.0    &   -9.1331(20)   &   -2.8358(20)  &   -11.538(6)\\
13.5    &   -9.3654(23)   &   -2.9111(16)  &   -11.822(6)\\
17.0    &   -9.7261(22)   &   -3.0181(16)  &   -12.276(6)\\
20.0    &   -9.9253(21)   &   -3.0862(17)  &   -12.525(6)\\
24.0    &  -10.1097(15)   &   -3.1420(17)  &   -12.768(7)\\
30.0    &  -10.2941(17)   &   -3.1987(18)  &   -12.995(6)\\
40.0    &  -10.4792(18)   &   -3.2587(6)   &   -13.240(6)\\
60.0    &  -10.6608(17)   &   -3.3148(7)   &   -13.446(7)\\
 \hline
\end{tabular}
\end{center}
\caption{Values of $\langle T^{[1]}_{0k} \rangle_{\vec\xi}$ measured at bare coupling
$\beta=6/g_0^2$ on lattices of size $48^3\times L_0/a$ and 
$\vec\xi=(1,0,0)$.\label{tab:t0k} }
\end{table}
%}

%\TABLE{
\begin{table}[htb]
\begin{center}
\begin{tabular}{| l | l | l |}
\hline
$\beta$ & $\langle T^{[1]}_{0k} \rangle_{\vec\xi} $ & $\langle T^{[3]}_{00}\rangle_{\vec\xi} - 
\langle T^{[3]}_{kk} \rangle_{\vec\xi} $ \\
& {\small $(\times 10^{-3})$} & {\small $(\times 10^{-3})$} \\
\hline
\multicolumn{3}{|c|}{$L/a=80 \qquad L_0/a = 4$}\\
\hline
6.3   & -4.1926(32)  & -5.920(7)     \\
7.5   & -5.2711(17)  & -7.228(5)   \\
10.0  & -6.1130(23)  & -8.155(6)     \\
15.0  & -6.7612(22)  & -8.825(6)     \\
24.0  & -7.2041(9)   & -9.2798(23)   \\
\hline
\multicolumn{3}{|c|}{$L/a=120 \qquad L_0/a = 6$}\\
\hline
6.3   & -0.7067(4)  & -1.0642(14)    \\
7.5   & -0.9703(5)  & -1.4242(14)    \\
10.0  & -1.1345(5)  & -1.6321(13)    \\
15.0  & -1.2559(5)  & -1.7761(14)    \\
24.0  & -1.3379(5)  & -1.8703(14)    \\
\hline
\multicolumn{3}{|c|}{$L/a=160 \qquad L_0/a = 8$}\\
\hline
6.3   & -0.18493(15)  &  -0.2839(6)     \\
7.5   & -0.29738(16)  &  -0.4475(5)   \\
10.0  & -0.35095(17)  &  -0.5190(5)     \\
15.0  & -0.38832(18)  &  -0.5666(5)     \\
24.0  & -0.41292(22)  &  -0.5972(6)     \\
\hline
\multicolumn{3}{|c|}{$L/a=240 \qquad L_0/a = 12$}\\
\hline
7.5   & -0.05643(11)  &  -0.08597(37)   \\
10.0  & -0.06785(12)  &  -0.10255(35)   \\
15.0  & -0.07494(16)  &  -0.1121(5)   \\
24.0  & -0.08029(17)  &  -0.1188(5)   \\
\hline
\end{tabular}
\end{center}
\caption{Values of $\langle T^{[1]}_{0k} \rangle_{\vec\xi}$ and $\langle T^{[3]}_{00} \rangle_{\vec\xi} - 
\langle T^{[3]}_{kk} \rangle_{\vec\xi}$
 measured at bare coupling $\beta=6/g_0^2$ with shift ${\vec\xi}=(1/2,0,0)$ and at
 fixed ratio $L/L_0=20$. \label{tab:ztbare}}
\end{table}
%}

%%%%%%%%%%%%%%%%%%%%%%%%%%%%%%%%%%%%%%%%%%%%%%%%%%%%%%%%%%%%%%%%
\bibliography{./lattice.bib}

\providecommand{\href}[2]{#2}\begingroup\raggedright\begin{thebibliography}{10}

\bibitem{Caracciolo:1989pt}
S.~Caracciolo, G.~Curci, P.~Menotti, and A.~Pelissetto, {\it {The energy
  momentum tensor for lattice gauge theories}},  {\em Annals Phys.} {\bf 197}
  (1990) 119.

\bibitem{Giusti:2010bb}
L.~Giusti and H.~B. Meyer, {\it {Thermal momentum distribution from path
  integrals with shifted boundary conditions}},  {\em Phys.Rev.Lett.} {\bf 106}
  (2011) 131601, [\href{http://xxx.lanl.gov/abs/1011.2727}{{\tt
  arXiv:1011.2727}}].

\bibitem{Giusti:2011kt}
L.~Giusti and H.~B. Meyer, {\it {Thermodynamic potentials from shifted boundary
  conditions: the scalar-field theory case}},  {\em JHEP} {\bf 1111} (2011)
  087, [\href{http://xxx.lanl.gov/abs/1110.3136}{{\tt arXiv:1110.3136}}].

\bibitem{Giusti:2012yj}
L.~Giusti and H.~B. Meyer, {\it {Implications of Poincare symmetry for thermal
  field theories in finite-volume}},  {\em JHEP} {\bf 1301} (2013) 140,
  [\href{http://xxx.lanl.gov/abs/1211.6669}{{\tt arXiv:1211.6669}}].

\bibitem{DellaMorte:2010yp}
M.~Della~Morte and L.~Giusti, {\it {A novel approach for computing glueball
  masses and matrix elements in Yang-Mills theories on the lattice}},  {\em
  JHEP} {\bf 1105} (2011) 056, [\href{http://xxx.lanl.gov/abs/1012.2562}{{\tt
  arXiv:1012.2562}}].

\bibitem{Suzuki:2013gza}
H.~Suzuki, {\it {Energy-momentum tensor from the Yang-Mills gradient flow}},
  {\em PTEP} {\bf 2013} (2013), no.~8 083B03,
  [\href{http://xxx.lanl.gov/abs/1304.0533}{{\tt arXiv:1304.0533}}].

\bibitem{DelDebbio:2013zaa}
L.~Del~Debbio, A.~Patella, and A.~Rago, {\it {Space-time symmetries and the
  Yang-Mills gradient flow}},  {\em JHEP} {\bf 1311} (2013) 212,
  [\href{http://xxx.lanl.gov/abs/1306.1173}{{\tt arXiv:1306.1173}}].

\bibitem{Makino:2014taa}
H.~Makino and H.~Suzuki, {\it {Lattice energy-momentum tensor from the
  Yang-Mills gradient flow--inclusion of fermion fields}},  {\em PTEP} {\bf
  2014} (2014), no.~6 063B02, [\href{http://xxx.lanl.gov/abs/1403.4772}{{\tt
  arXiv:1403.4772}}].

\bibitem{Luscher:2013lgggg}
M.~L{\"uscher}, {\it {Small flow-time expansion of gauge-invariant local
  fields}},  {\em Unpublished notes June 2013}.

\bibitem{Adler:1976zt}
S.~L. Adler, J.~C. Collins, and A.~Duncan, {\it {Energy-Momentum-Tensor Trace
  Anomaly in Spin 1/2 Quantum Electrodynamics}},  {\em Phys.Rev.} {\bf D15}
  (1977) 1712.

\bibitem{Collins:1976yq}
J.~C. Collins, A.~Duncan, and S.~D. Joglekar, {\it {Trace and Dilatation
  Anomalies in Gauge Theories}},  {\em Phys.Rev.} {\bf D16} (1977) 438--449.

\bibitem{Caracciolo:1991cp}
S.~Caracciolo, P.~Menotti, and A.~Pelissetto, {\it {One loop analytic
  computation of the energy momentum tensor for lattice gauge theories}},  {\em
  Nucl.Phys.} {\bf B375} (1992) 195--242.

\bibitem{Giusti:2014ila}
L.~Giusti and M.~Pepe, {\it {Equation of state of a relativistic theory from a
  moving frame}},  {\em Phys.Rev.Lett.} {\bf 113} (2014) 031601,
  [\href{http://xxx.lanl.gov/abs/1403.0360}{{\tt arXiv:1403.0360}}].

\bibitem{Robaina:2013zmb}
D.~Robaina and H.~B. Meyer, {\it {Renormalization of the momentum density on
  the lattice using shifted boundary conditions}},  {\em PoS} {\bf LATTICE2013}
  (2014) 323, [\href{http://xxx.lanl.gov/abs/1310.6075}{{\tt
  arXiv:1310.6075}}].

\bibitem{Cabibbo:1982zn}
N.~Cabibbo and E.~Marinari, {\it {A New Method for Updating SU(N) Matrices in
  Computer Simulations of Gauge Theories}},  {\em Phys. Lett.} {\bf B119}
  (1982) 387--390.

\bibitem{deForcrand:2000fi}
P.~de~Forcrand, M.~D'Elia, and M.~Pepe, {\it {A study of the 't Hooft loop in
  SU(2) Yang-Mills theory}},  {\em Phys. Rev. Lett.} {\bf 86} (2001) 1438,
  [\href{http://xxx.lanl.gov/abs/hep-lat/0007034}{{\tt hep-lat/0007034}}].

\bibitem{DellaMorte:2007zz}
M.~Della~Morte and L.~Giusti, {\it {Exploiting symmetries for exponential error
  reduction in path integral Monte Carlo}},  {\em Comput. Phys. Commun.} {\bf
  180} (2009) 813--818.

\bibitem{Giusti:2014tfa}
  L.~Giusti and M.~Pepe,
  {\it Non-perturbative renormalization of the energy-momentum tensor in SU(3) Yang-Mills theory}, {\em Proc. Sci.} {\bf LATTICE2014} (2015) 322

\bibitem{Becchi:1974xu}
C.~Becchi, A.~Rouet, and R.~Stora, {\it {The Abelian Higgs-Kibble Model.
  Unitarity of the S Operator}},  {\em Phys.Lett.} {\bf B52} (1974) 344.

\bibitem{Becchi:1974md}
C.~Becchi, A.~Rouet, and R.~Stora, {\it {Renormalization of the Abelian
  Higgs-Kibble Model}},  {\em Commun.Math.Phys.} {\bf 42} (1975) 127--162.

\bibitem{Tyutin:1975qk}
I.~Tyutin, {\it {Gauge Invariance in Field Theory and Statistical Physics in
  Operator Formalism}},  \href{http://xxx.lanl.gov/abs/0812.0580}{{\tt
  arXiv:0812.0580}}.

\bibitem{Callan:1970ze}
J.~Callan, Curtis~G., S.~R. Coleman, and R.~Jackiw, {\it {A New improved
  energy-momentum tensor}},  {\em Annals Phys.} {\bf 59} (1970) 42--73.

\bibitem{Weisz:2010nr}
P.~Weisz, {\it {Renormalization and lattice artifacts}},
  \href{http://xxx.lanl.gov/abs/1004.3462}{{\tt arXiv:1004.3462}}.

\end{thebibliography}\endgroup
%%%%%%%%%%%%%%%%%%%%%%%%%%%%%%%%%%%%%%%%%%%%%%%%%%%%%%%%%%%%%%%%

\end{document}